\documentclass[twocolumn,showpacs,preprintnumbers,pra]{revtex4}

\usepackage{graphicx}
\usepackage{amsmath}
\usepackage{amssymb}

\newcommand{\e}{\text{e}}
\newcommand{\ii}{\text{i}}
\newcommand{\dd}{\text{d}}
\DeclareMathOperator{\F}{F}

\begin{document}

\title{Laser driven atoms in half-cavities}

\author{U. Dorner and P. Zoller} \affiliation{Institute for
  Theoretical Physics, Technikerstra\ss e 25, A-6020 Innsbruck, Austria}
        
\date{\today}

\begin{abstract}
  The behavior of a two level atom in a half-cavity, i.e.\ a cavity
  with one mirror, is studied within the framework of a one
  dimensional model with respect to spontaneous decay and resonance
  fluorescence. The system under consideration corresponds to the
  setup of a recently performed experiment [J. Eschner \textit{et.
    al.}, Nature \textbf{413}, 495 (2001)] where the influence of a
  mirror on a fluorescing single atom was revealed.  In the present
  work special attention is paid to a regime of large atom-mirror
  distances where intrinsic memory effects cannot be neglected
  anymore.  This is done with the help of delay differential equations
  which contain, for small atom-mirror distances, the Markovian limit
  with effective level shifts and decay rates leading to the
  phenomenon of enhancement or inhibition of spontaneous decay.
  Several features are recovered beyond an effective Markovian
  treatment, appearing in experimental accessible quantities like
  intensity or emission spectra of the scattered light.
\end{abstract}

\pacs{42.50.Ct, 42.50.Md, 32.80.-t, 32.70.Jz}

\maketitle

\section{\label{sec:sec1}Introduction}
The change in the behavior of atoms when the structure of the
``surrounding'' field differs from that of free space is treated so
far in innumerable works and is essentially the basic topic of cavity
QED \cite{berman,hinds,haroche}.  Effects like modified decay rates of
atoms in cavities were visible in various measurements
\cite{heinzen87a,heinzen87b,jhe87,hulet85,goy83,martini}.  In this
case, the atoms couple irreversibly to a large number of field modes
and the problem can be treated within the framework of perturbation
theory. This regime is therefore known as ``low-Q'' or perturbative
cavity QED \cite{hinds1994}. Another area of high importance is its
counterpart, the physics of ``high-Q'' cavities, where the atoms
interact strongly only with one (or few) field mode(s).  In this
context, recent experiments include, e.g., the observation of atom
trajectories in cavities storing merely one photon
\cite{kimble,rempe}.  Furthermore, among other things, effects caused
by the spatial structure of a field mode in a cavity were demonstrated
\cite{blatt2002,walther2001}.  High-Q cavities serve also as a testing
ground for fundamental quantum mechanical effects like entanglement or
decoherence \cite{haroche2001,walther1994}.

Besides some considerations on the spontaneous decay of an excited
two-level atom we will mainly focus in this paper on the problem of
resonance fluorescence in a half-cavity, i.e.\ a cavity with one
mirror, where we pay special attention to the position dependence of
the atomic dynamics.  To this end we will particularly consider a
physical system which essentially coincides with the setup of an
recently performed experiment \cite{eschner}.  Here, the radiation
which is emitted by a laser cooled ion stored in Paul trap is partly
collimated by a lens and reflected back by a mirror to the atom. The
intensity of the scattered light was measured as a function of the
mirror position leading to an oscillatory behavior of the photon
counting rate, proving the existence of inhibited and enhanced
spontaneous emission effects.  In this case, where the atom is
relatively close to the mirror, the observed effects can in principal
explained by introducing some effective modified (position dependent)
spontaneous emission rates and level shifts.  This could be done since
the time the light needs to bounce back and forth between the atom and
the mirror could be set essentially to zero (Markovian limit). The
situation is more complicated when the distance between the atom and
the mirror is large.
  
In this paper we will, among other things, particularly consider this
case and it turns out that the dynamics of the atom can be described
generally in terms of non-Markovian (delay-differential) equations.
As we will see, the distance between the atom and the mirror
influences the atomic behavior essentially on two scales: On the one
hand, on a large scale, i.e.\ whether the atom is located far away from
the mirror or very close to the mirror. This scale can be measured
essentially by a dimensionless quantity $\Gamma_0\tau$ where
$\Gamma_0$ is given by the width of the field spectrum (in case of
vanishing laser intensity it is simply the atomic spontaneous emission
rate) and the time $\tau$ the light needs for a round trip between
atom and mirror.  On the other hand, the atomic behavior varies also
if the distance is changed on the scale of an optical wavelength given
by $\omega_L\tau$.  For example in case of a small atom-mirror
distance ($\Gamma_0\tau\ll1$) the equations of motions become
approximately Markovian and the well-known phenomenon of enhanced or
inhibited spontaneous emission (depending on $\omega_L\tau$), can be
recovered. Thus, it is possible to describe the system by introducing
effective spontaneous emission rates and level shifts.  In general,
however, the retardation of the time argument in the equations of
motion cannot be neglected.  We will not consider in this paper
effects arising in the case of extremely small distances, i.e.\ of the
order of wavelengths or smaller, between the atom and the mirror
\cite{milonni1994,hinds,hinds1994,meschede,sandoghar}.

In connection with cavity QED, in the broadest sense, the above mentioned
delay-differential equations appeared already in some works.  These
include for example the analytical treatments of Milonni \textit{et al.}
\cite{acker} in which a single excited quantum system coupled to an
infinite set of equally spaced discrete levels was considered, a
system which reminds of an atom in a cavity but without taking into
account some position depending effects. The latter problem was
discussed in the framework of a one-dimensional model by Cook and
Milonni \cite{cook} in case of an excited atom in front of a partially
reflecting wall (modeled as a collection of two-level atoms) and in a
Fabry-Perot resonator. This treatment is closely related to our
discussion of pure spontaneous decay since we will recover the same
equation of motion.  Another treatment of this problem was given by
Feng and Ujihara \cite{ujihara90} by using an appropriate set of mode
functions in order to account for non-perfect mirror reflectivities.
Dung and Ujihara \cite{ujihara99} finally examined an atom in a
three-dimensional Fabry-Perot resonator.  Although a delay
differential equation was not explicitly formulated, retardation
effects in the interaction of two atoms were also discussed in
\cite{knight74,knight75}. 
A partly numerical examination of an atom inside a spherical cavity
was given by Parker and Stroud \cite{parker}.  Numerical examinations
include furthermore the work of Gie{\ss}en \textit{et al.}\cite{giessen}
and Bu{\v{z}}ek \textit{et al.}  \cite{buzek} both treating an atom in a
(one-dimensional) cavity whereas the latter also investigates the
presence of material media.  The mentioned works have in common that
recurrences of the atomic population take place for large dimensions
of the cavity which is due to one-photon wave packets bouncing back
and forth between the cavity walls.  However, there is always only one
excitation contained in the system making the problem accessible for
analytical considerations.

In case of a (near-)resonantly driven atom the dynamics of the system
is more complicated since the number of excitations increase
continuously.  The scattered radiation will be reflected by the mirror
and re-interact with its own source, the atom. This situation reminds
us of a feedback problem \cite{wiseman1994a,wiseman1994b} where mostly
the assumption of a negligible feedback time delay is made. However, the
situation of large atom-mirror distances would correspond to a
non-Markovian feedback \cite{giovannetti,wang}.

This article is structured as follows. In Sec.~\ref{sec:sec2} we will
reconsider the problem of pure spontaneous decay of an excited atom in
the presence of a mirror while in Sec.~\ref{sec:sec3} a continuous
laser excitation of the atom is incorporated to our examination. We
will discuss several limits including low and higher laser intensities
and small distances between the atom and the mirror.  Finally, a
summary is given in Sec.~\ref{sec:summary} and cumbersome formulas and
calculations are moved to appendices.

\section{\label{sec:sec2}Spontaneous emission}
In this section we will investigate the spontaneous emission of an
atom at rest in the presence of a mirror initially prepared in the excited state.
We will derive a non-Markovian equation of motion for this system. 
Although this derivation is related to the calculations in \cite{cook}
it will be discussed here, not only to introduce our notation but also
to present new, entirely analytical results, also with respect to
spectral properties of the emitted light.  Furthermore it will turn
out that some calculation methods can be transfered and some features
of this system are recovered when we include laser excitation in our
considerations.
\subsection{The model\label{sec:model}}
\begin{figure}[t]
  \centering \includegraphics[]{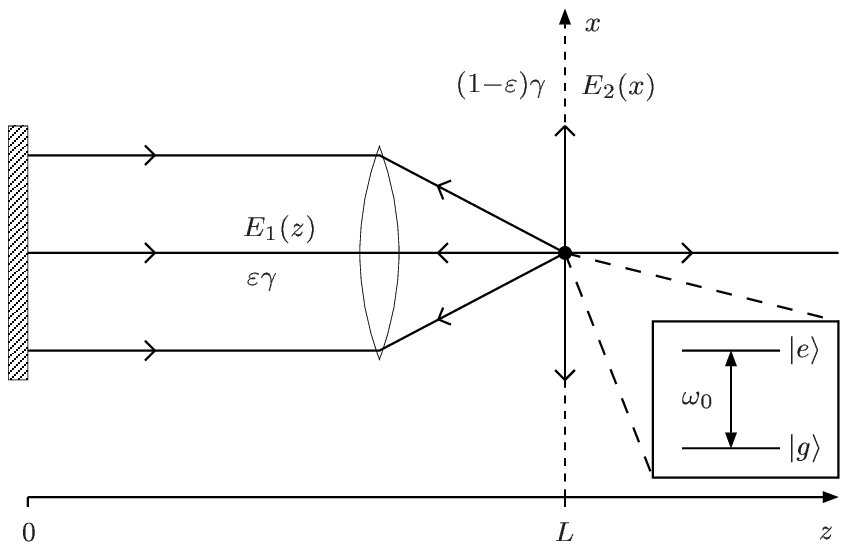}
 \caption{Sketch of the physical system under consideration. 
   The radiation emitted by a two-level atom is partly reflected (via
   a lens) back to the atom which is modeled by an atom coupled to two
   one-dimensional channels with a different mode structure (running
   and standing wave modes, respectively). \label{fig1}}
\end{figure}

As already mentioned we examine an initially excited two-level atom
with transition frequency $\omega_0$ in the presence of a finite size
mirror where the light emitted in a certain solid angle fraction
$\varepsilon$ is reflected back to the atom.  This is
achieved by a lens which collimates the radiation before it is
reflected \cite{eschner}.  The remaining emission is not affected by
the mirror. Thus, it is reasonable to consider the coupling of the atom
to two reservoirs (or ``channels'') consisting of one-dimensional
fields with standing wave field modes and running wave field modes,
respectively (see Fig.~\ref{fig1}), i.e.\ the Hamiltonian in rotating
wave approximation reads
\begin{eqnarray}
H = H_0 &-& d\big(E^\dagger_1(L)\sigma_- + \sigma_+E_1(L)\big) \nonumber\\
        &-& d\big(E^\dagger_2(0)\sigma_- + \sigma_+E_2(0)\big)
\end{eqnarray}
with
\begin{eqnarray}
&&H_0 =  \hbar\omega_0\sigma_+\sigma_- + \int \dd k\,\hbar\omega_k a_k^\dagger a_k 
                                       + \int \dd k\,\hbar\omega_k b_k^\dagger b_k,\nonumber\\
&&E_1(z) = \ii\int \dd k\,\alpha_k\sin(kz)a_k,\quad k>0, \nonumber\\
&&E_2(x) = \ii\int \dd k\,\beta_k\e^{\ii kx} b_k,\quad k\in\mathbb{R},
\label{eq:field_op}
\end{eqnarray}
and $\omega_k=\vert k\vert c$.  In contrast to a cavity, here, the
mode density of the mirror channel is continuous since only one
boundary condition has to be fulfilled.
The operators $\sigma_+,\,\sigma_-$ are the usual raising and lowering
operators of a two-level system with upper level $\vert e\rangle$ and ground
state $\vert g\rangle$, $\sigma_+ = \vert e\rangle\langle g\vert$, $\sigma_- =
\vert g\rangle\langle e\vert$, and $a_k^\dagger,\,b_k^\dagger,\,a_k,\,b_k$ are
creation and annihilation operators of a photon in the $k$th mode of
the different environments.  The dipole matrix element $d$ is assumed
to be real and for the sake of simplicity we suppress the vectorial
character of $d$ and $E$. The exact form of the factors $\alpha_k$ and
$\beta_k$ is of no importance here, we merely assume that they are
approximately constant in a frequency range of relevance (usually they
have a frequency dependence $\alpha_k\sim\sqrt{\omega_k}$ and
$\beta_k\sim\sqrt{\nu_k}$). In order to investigate the dynamics of
the system we make the Wigner-Weisskopf type ansatz
\begin{eqnarray}
\vert\psi(t)\rangle = &&b_e(t)\vert e,\{0\}_1,\{0\}_2\rangle \nonumber\\ 
&+& \int \dd k\,b_{g,k}^1(t)\vert g,\{k\}_1,\{0\}_2\rangle \nonumber\\
&+& \int \dd k\,b_{g,k}^2(t)\vert g,\{0\}_1,\{k\}_2\rangle,
\label{eq:ansatz}
\end{eqnarray}
where $\vert\{0\}\rangle$ denotes the vacuum state of the radiation field
and $\vert\{k\}\rangle$ the state with exactly one photon in mode $k$.  We
will consider here an initially excited atom in the absence of any
photon, i.e.\ $b_e(0)=1$ and $b_{g,k}^j(0)=0$.  In contrast to the
notation in Eq.~(\ref{eq:ansatz}), in the following the amplitudes are
always taken in a rotating frame, i.e.\ we make the substitutions
$b_e(t) \rightarrow b_e(t)\e^{-\ii\omega_0t}$ and $b_{g,k}^j(t)
\rightarrow b_{g,k}^j(t)\e^{-\ii\omega_kt}$ .
With the help of the ``essential states'' contained in the above
ansatz it is possible to write down a closed set of equations of
motion for the amplitudes which take the form
\begin{subequations}
\label{eq:dgl0}
\begin{eqnarray}
&&\dot b_e(t) 
= -\int \dd k\,g_k\sin(kL)
           \e^{-\ii(\omega_k - \omega_0)t}b_{g,k}^1(t) \nonumber\\
&& \qquad\quad -\int \dd k\, h_k
           \e^{-\ii(\omega_k - \omega_0)t}b_{g,k}^2(t)\label{eq:dgl_a},\\
&&\dot b_{g,k}^1(t) = g_k\sin(kL)
           \e^{\ii(\omega_k - \omega_0)t}b_e(t)\label{eq:dgl_b},\\
&&\dot b_{g,k}^2(t) = h_k
           \e^{\ii(\omega_k - \omega_0)t}b_e(t)\label{eq:dgl_c}
\end{eqnarray}
\end{subequations}
with $g_k\equiv\alpha_kd/\hbar$ and $h_k\equiv\beta_kd/\hbar$.

By formally integrating the latter two equations and inserting them
into the first one we get
\begin{eqnarray}
\dot b_e(t) 
&=& -\int_0^t\dd t'\,b_e(t')\int \dd k\,
          g_k^2\sin^2(kL)\e^{\ii(\omega_k - \omega_0)(t' - t)} \nonumber\\
  &&-\int_0^t\dd t'\,b_e(t')\int \dd k\,
          h_k^2\e^{\ii(\omega_k - \omega_0)(t' - t)} \nonumber\\
&=& - \gamma \int_0^t \dd t\,b_e(t')\delta(t' - t) \e^{-\ii\omega_0(t'-t)}\nonumber\\
&& +  \varepsilon\frac{\gamma}{2}\int_0^t\dd t'\,b_e(t')
      [\delta(t'-t+\tau) \nonumber\\ 
&&\hspace{0.25\columnwidth} + \delta(t'-t-\tau)]\e^{-\ii\omega_0(t'-t)}.
\label{eq:ww_approx}
\end{eqnarray}
Here, we introduced the free space spontaneous decay rate $\gamma$
which is split up into a part $\varepsilon\gamma\equiv\pi g_{k_0}^2/c$
and $(1-\varepsilon)\gamma\equiv4\pi h_{k_0}^2/c$, corresponding to the
coupling of the atom to the first and second channel.  The quantity
$\varepsilon$ is the solid angle fraction which is covered by the lens
since it characterizes the fraction of radiation which is reflected.
In the above equations we also introduced the time $\tau\equiv 2L/c$
the light needs for the distance atom-mirror-atom.  Furthermore, in
the first step of (\ref{eq:ww_approx}), a Wigner-Weisskopf type
approximation was made based on the well-known fact that the relative
variation of $g_k^2,\,h_k^2$($\sim\omega$) is very slow in the domain
where the double integration in the first line of (\ref{eq:ww_approx})
lead to appreciable values. Diverging terms connected with level shifts
are omitted.  It should be stressed here that in this paper only
distances between the atom and the mirror are considered which are
much larger than an optical wavelength, i.e.\ $\omega_0\tau\gg1$ which
is in an optical frequency domain already the case for, say, a
millimeter.  

Eq.~(\ref{eq:ww_approx}) yields finally the delay
differential equation
\begin{equation}
\label{eq:delay1}
\dot b_e(t) = -\frac{\gamma}{2}b_e(t) 
  + \varepsilon\frac{\gamma}{2}\e^{\ii\omega_0\tau} b_e(t - \tau)\Theta(t - \tau),
\end{equation}
where $\Theta(t)$ is the Heaviside step function.  The first term on
the right hand side of this equation corresponds to the usual free
space exponential decay while the second term represents the effect of
the reflected radiation on the atom which was emitted at time
$\tau$ \textit{before} it interacts again with the atom.  Thus, the
retarded argument of the excited state amplitude directly indicates
the memory effects which are inherent in the system.  Furthermore, the
second term is weighted with the factor $\varepsilon$ revealing that
only a fraction of the emitted light is reflected.

Eq.~(\ref{eq:delay1}) is a delay-differential equation~\cite{driver}
and since we will encounter in Sec.~\ref{sec:low_laser} and
Sec.~\ref{sec:high_laser} some more complicated equations of this type
we move a further discussion of some general properties of equations
of this kind to these sections.

\subsection{\label{sec:sec2a}Discussion}
Using Laplace transformation and geometric series expansion
Eq.~(\ref{eq:delay1}) can easily be solved and one obtains
\begin{equation}
\label{eq:sol1}
b_e(t) = \sum_{n=0}^{\infty}\frac{(\varepsilon\gamma/2)^n}{n!}\e^{\ii\omega_0n\tau}
                        \e^{-\frac{\gamma}{2}(t-n\tau)}(t-n\tau)^n\Theta(t-n\tau).
\end{equation}
It should be mentioned that this expression can also be
obtained by a direct Laplace transformation of the Schr\"odinger
equation (\ref{eq:dgl0}) \cite{knight74}.  

The above solution reveals that the systems dynamics has a ``step''
character which can be seen most easily if one divides the time axis
into intervals of length $\tau$. For $t\in[0,\tau]$ the sum consists
only of one term, $\exp(-\gamma\tau/2)$, which coincides with the free
space behavior of a decaying atom. The physical reason for that is
that the atom requires at least the time the light needs to get from the
atom to the mirror and back to the atom again in order to ``see'' the
mirror. For $t\in[\tau,2\tau]$ the amplitude consists of two terms,
\begin{equation}
b_e(t) = \e^{-\frac{\gamma}{2}t} 
       + \varepsilon\frac{\gamma}{2}\e^{\ii\omega_0\tau}
                    \e^{-\frac{\gamma}{2}(t-\tau)}(t-\tau),
\end{equation}
giving rise to an interference term in the probability of finding the
atom in the excited state. The second term is due to the emitted
radiation reflected back to the atom.  The light the atom emits right
now arrives the atom again at the beginning of the third time interval
where the sum in~(\ref{eq:sol1}) includes a further term and so on.

\begin{figure}[t]
  \centering \includegraphics[]{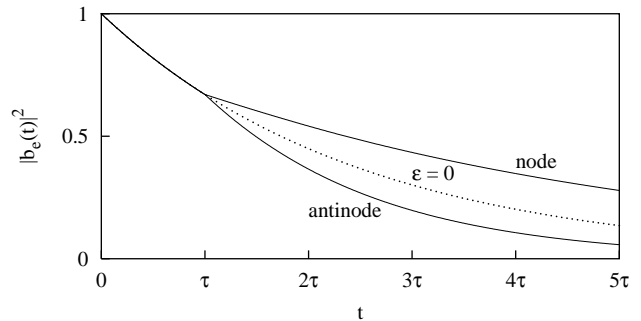}
 \caption{
   Upper state population of an atom close to the mirror for different
   exact positions, i.e.\ $\omega_0\tau=2n\pi$ (node) and
   $\omega_0\tau=(2n+1)\pi$ (antinode).  The remaining parameters are
   $\varepsilon = 0.4$, $\gamma\tau = 0.4$.  Also indicated is the
   corresponding free space solution ($\varepsilon=0$).\label{fig2}}
\end{figure}

The role of the interference terms in the excited state probabilities
strongly depends on the distance between the atom and the mirror which
can be measured by the quantity $\gamma\tau$. If we consider again the
second time interval it is easy to see that that the interference term
is of order $\gamma\tau$ while a further term is of order
$(\gamma\tau)^2$ which can be neglected for $\gamma\tau\ll1$ (small
distance). Hence we get the expression
\begin{equation}
\vert b_e(t)\vert^2 \approx \e^{-\gamma t}
(1 + \varepsilon\gamma(t-\tau)\cos(\omega_0\tau)),\quad t\in[\tau,2\tau],
\end{equation} 
where we guess the beginning of an exponential series.  The
examination of the dynamics in this limit for larger times based on
Eq.(\ref{eq:sol1}) is relatively complicated. It is more convenient to
return to the delay differential equation~(\ref{eq:delay1}). Since we
are working in a rotating frame the amplitude $b_e(t)$ varies slowly
on a time scale given by $1/\gamma$. Thus, in the limit
$\gamma\tau\ll1$ we can make the approximation $\tau\rightarrow +0$ in
the argument of $b_e$ in the second term on the right hand side of
Eq.~(\ref{eq:delay1}) and obtain the Markovian equation
\begin{equation}
\label{eq:dgl1}
\dot b_e(t) \approx
 \begin{cases} -\frac{\gamma}{2}b_e(t) &;\,t\le\tau \\
   -\frac{\gamma}{2}(1-\varepsilon\e^{\ii\omega_0\tau})b_e(t)
   &;\,t>\tau.
 \end{cases}
\end{equation}
This leads to the excited state probability
\begin{align}
\label{eq:sol1_app1}
&\vert b_e(t)\vert^2 \approx
 \begin{cases} 
   \e^{-\gamma t}&;\,t\le\tau \\
   \e^{-\gamma\tau}\e^{-\tilde\gamma(t-\tau)}&;\,t>\tau
 \end{cases} 
\end{align}
with $\tilde\gamma = \gamma(1-\varepsilon\cos(\omega_0\tau))$.  The
upper state population based on the exact amplitude~(\ref{eq:sol1}) in
this limit is shown in Fig.~\ref{fig2}.  The behavior of the curves
coincide almost perfectly with the predictions of
Eq.~(\ref{eq:sol1_app1}): After a period of length $\tau$ there is an
enhancement or inhibition of spontaneous decay depending on the factor
$1-\varepsilon\cos(\omega_0\tau)$ which corresponds to the amplitude
of a standing wave mode $\sin{(k_0z)}$ at the position of the atom: In a
node of the standing wave, spontaneous decay is inhibited while in an
antinode it is enhanced.

\begin{figure}[t]
  \centering \includegraphics[]{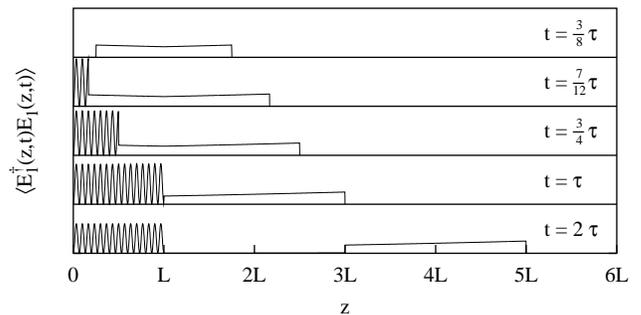}
 \caption{
   Intensity of the field at different points of time for the
   parameters of Fig.~\ref{fig2}.  The atom is located at a node (i.e.\
   $\omega_0\tau=2n\pi$) at $z=L$.\label{fig3}}
\end{figure}

\begin{figure}[t]
  \centering \includegraphics[]{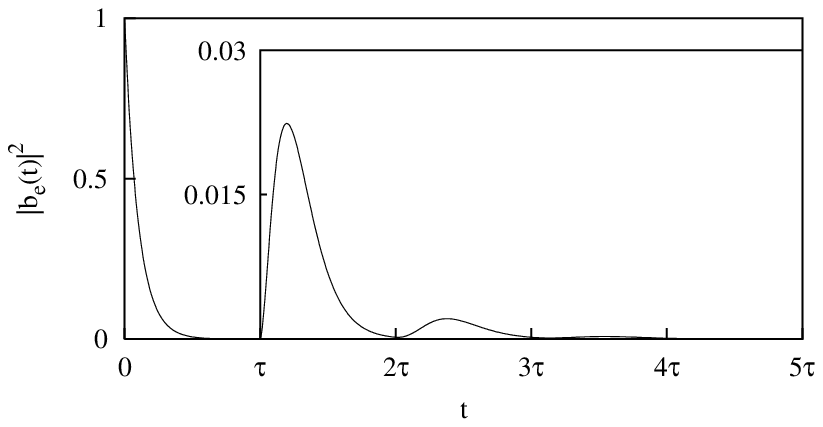}
 \caption{
   Upper state population of an atom far away from the mirror and
   $\omega_0\tau=2n\pi$.  The remaining parameters are $\varepsilon =
   0.4$, $\gamma\tau = 10$. The inset is a vertical
   magnification.
\label{fig4}}
\centering \includegraphics[]{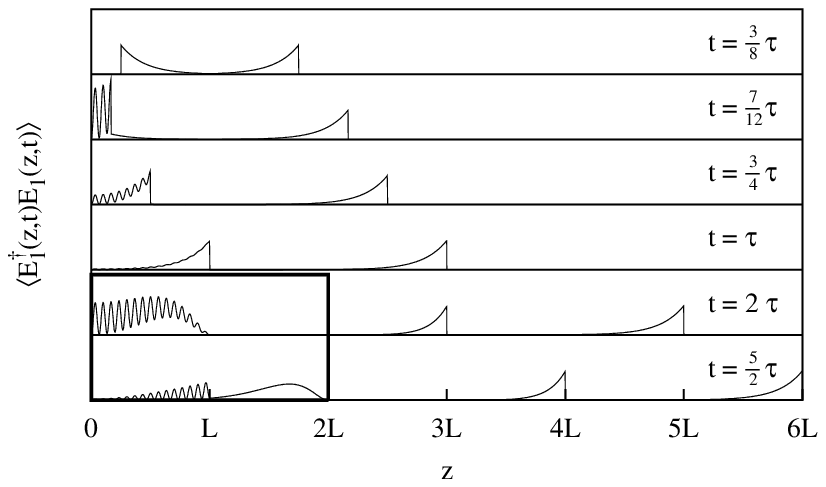}
 \caption{
   Intensity of the field at different points of time for the
   parameters of Fig.~\ref{fig4}.  The atom is located again at $z=L$
   while the inset is a vertical magnification.\label{fig5}}
\end{figure}

To get a more physical insight in this behavior the intensity of the
electric field $E_1$ which is reflected by the mirror depending on
space and time is shown in Fig.~\ref{fig3}. The derivation of an
analytical expression for $\langle E_1^\dagger(z,t) E_1(z,t)\rangle$
can be found in Appendix~\ref{app:1}. In a realistic situation with
regard to the setup considered here the frequency of the oscillations
which can be seen in Fig.~\ref{fig3} (and also in Fig.~\ref{fig5})
would be significantly higher than indicated in these figures.
However, for the sake of visibility, a rather small frequency is
chosen here.
Due to the small distance between the atom (located at $z=L$) and the
mirror (located at $z=0$) the reflected light has the possibility to
interfere with the radiation which is still emitted by the atom
leading to a standing wave pattern of the form $\sim\sin^2(k_0z)$ that
has, in case of the example shown in Fig.~\ref{fig3}, a node at the
position of the atom, i.e.\ a zero electric field. Due to this fact,
further emission of radiation in this channel is suppressed.  Another
interesting feature with respect to Fig.~\ref{fig3} is that the
amplitude of the standing wave decreases for $\varepsilon<1$ with
increasing time whereas the energy escapes in the other channel. The
situation is reminiscent of a cavity where the atom acts like a
partially transmitting mirror.

In the limit of large distances between atom and mirror (i.e.\
$\gamma\tau\gg1$) the sum~(\ref{eq:sol1}) is dominated by the term
with the highest power of $\gamma\tau$. Hence, we get in a time
interval $[m\tau,(m+1)\tau]$
\begin{equation}
\vert b_e(t)\vert^2 \approx 
 \left(\frac{(\varepsilon\gamma/2)^m}{m!}\right)^2
              \e^{-\gamma(t-m\tau)}(t-m\tau)^{2m}.
\end{equation}
We see that the atom is partially reexcited by the radiation which it
has emitted before and that the exact position (node or antinode) is
not significant. This is illustrated in Fig.~\ref{fig4} and
Fig.~\ref{fig5} where we plot again the exact solution for the
excited state amplitude and the field intensity.
The atom is placed in a node of the standing wave and the inset
in Fig.~\ref{fig5} is a magnification in vertical direction of that
part. We see that interference between outgoing and incoming radiation
is much weaker than in Fig.~\ref{fig3}.

\begin{figure}[t]
  \centering \includegraphics[]{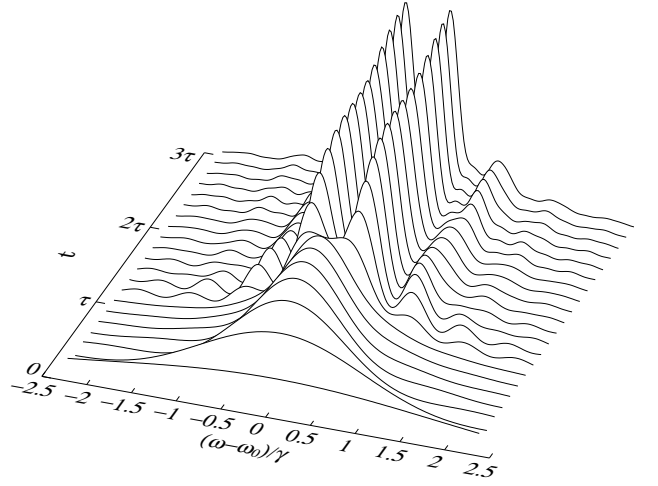}
 \caption{
   Transient photon population in channel $2$ depending on frequency
   and time for a large distance between the atom (placed at an
   antinode, i.e.\  $\omega_0\tau = (2n+1)\pi$) and the mirror.
   Further parameters are $\varepsilon = 0.4$ and $\gamma\tau = 10$.
\label{fig6}}
\end{figure}

\begin{figure}[t]
  \centering \includegraphics[]{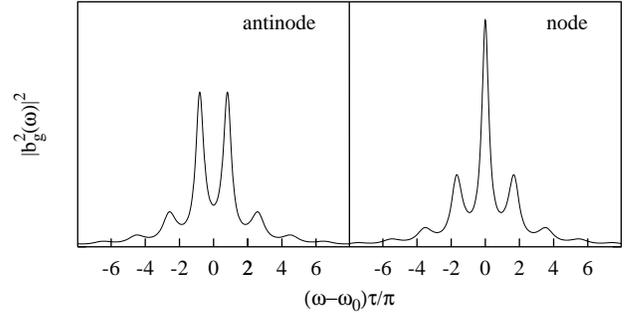}
 \caption{
   Frequency dependent steady state photon population in case of a
   large distance between the atom and the mirror.  The exact position
   of the atom is a node ($\omega_0\tau = 2n\pi$) or an antinode
   ($\omega_0\tau = (2n+1)\pi$), respectively ($\varepsilon = 0.4$,
   $\gamma\tau = 10$).
 \label{fig7}}
\end{figure}

In this context it is also interesting to take a look at the spectrum
of the emitted light (here this means the probability of finding a
photon of frequency $\omega$ in the long time limit). This can easily
be calculated by integrating Eq.~(\ref{eq:dgl_b}) and
Eq.~(\ref{eq:dgl_c}) and using Eq.~(\ref{eq:sol1}) which leads to
\begin{eqnarray}
b_g^j(\omega,t) &=& \frac{A_j(\omega)}{\frac{\gamma}{2}+\ii(\omega_0-\omega)}
  \sum_{n=0}^\infty\frac{(\varepsilon\gamma/2)^n}{n!}\e^{\ii\omega n\tau}(t-n\tau)^n
\nonumber\\
&&\quad \times G_n[-({\textstyle\frac{\gamma}{2}} 
                          + \ii(\omega_0-\omega))(t-n\tau)]\Theta(t-n\tau) \nonumber\\
\label{eq:trans_spec}
\end{eqnarray}
with
\begin{equation}
G_n[s] \equiv \sideset{_1}{_1}\F[n,n+1;s] - \e^{s} , 
\label{eq:Gn}
\end{equation}
where $\sideset{_1}{_1}\F[n,m;x]$ is the confluent hypergeometric
function and
\begin{equation}
A_j(\omega) \equiv 
\begin{cases}
  \sqrt{\frac{\varepsilon\gamma}{\pi}}\sin(\omega\tau/2),\quad j=1 \\
  \sqrt{\frac{(1-\varepsilon)\gamma}{2\pi}},\quad j=2.
\end{cases}
\end{equation}
The transient photon population of the second channel in case of a
relatively large atom-mirror distance for an atom placed in an
antinode of the resonant standing wave field mode is shown in
Fig.~\ref{fig6}.
In the long time limit the spectra take the form
\begin{equation}
\label{eq:spec1}
\vert b_{g}^j(\omega)\vert^2 = \frac{A_j^2(\omega)}
      {\frac{\gamma^2}{4}(1-\varepsilon\cos(\omega\tau))^2 
    + \left(\varepsilon\frac{\gamma}{2}\sin(\omega\tau) + \omega - \omega_0\right)^2}.
\end{equation}
Fig.~\ref{fig7} shows the steady state photon population of the
channel parallel to the mirror for different atomic positions.
The frequencies of the local minima in Fig.~\ref{fig6} and
Fig.~\ref{fig7} coincide approximately with the frequencies of those
standing wave modes which have an antinode at the position of the
atom.  This means that the photon distribution in the second
channel, which has initially roughly the shape of a Lorentzian (at the
end of the time interval $[0,\tau]$, see Fig.~\ref{fig6}), is
affected by the back reflected light in the first channel. Here, 
the radiation components with the mentioned frequencies have a higher
probability to be reabsorbed and emitted again (perhaps in modes of
other frequencies). This leads to a lower population of these modes.

If the atom is very close to the mirror we recognize that the differential equation
Eq.~(\ref{eq:dgl1}) contains the complex phase $\e^{\ii\omega_0\tau}$,
where the imaginary part of this factor can be interpreted as a level
shift.  This has consequences for the spectrum which takes in this
limit the form
\begin{equation}
\label{eq:spec1_shift}
\vert b_{g}^2(\omega)\vert^2 \sim \frac{1}
                       {\tilde\gamma^2/4 + (\omega - \tilde\omega_0)^2},
\end{equation}
where $\tilde\gamma$ is defined in Eq.~(\ref{eq:dgl1}) and
\begin{equation}
\tilde\omega_0\equiv\omega_0-\varepsilon\frac{\gamma}{2}\sin(\omega_0\tau).
\label{eq:shift}
\end{equation}
This expression can be derived with the help of Eq.~(\ref{eq:dgl1})
(where we neglect the small contribution arising from the first time
interval $[0,\tau]$) or with Eq.~(\ref{eq:spec1}) using the fact that
for $\gamma\tau\ll1$ the trigonometric functions in the denominator of
Eq.~(\ref{eq:spec1}) vary very slowly on a frequency scale $\gamma$.
The form of this spectrum illustrates first of all again the behavior
shown in Fig.~\ref{fig2}, i.e.\ the width of the Lorentzian is larger
or smaller depending whether the atom is placed in an antinode or a
node of a standing wave $\sin(k_0z)$. On the other hand the maximum of
the function is shifted according to the imaginary part of the
mentioned phase.  The physical interpretation of this shift is based
on the fact that the atom interacts with its own radiation.  It
corresponds to the energy of the atomic dipole in the reflected
electric field \cite{hinds1994,hinds,meschede,sandoghar}.
 
\section{\label{sec:sec3}Laser excitation}
The system discussed in the previous section was amenable to an
exact analytical treatment since the equations of motion decoupled by
using a Wigner-Weisskopf (or Markov)-type approximation. This reduced
the problem essentially to the solution of one equation describing
only the atomic dynamics. This was possible because the system
contained merely one excitation. The dynamics of the atom-field system
becomes more complicated when we include a continuous laser
excitation of the atom. The number of photons scattered by the atom
into the two channels will permanently increase and a part of them
will be reflected back, interacting again with the atom in addition to
the laser light. The atom starts now to emit a different kind of radiation which again returns to
the atom after some time and so on. Thus we expect that an electric
field is constituted with a complex structure.
The behavior of the system reminds us of that of a cascaded quantum
system
\cite{carmichael93,gardiner93,carmichael94,gardiner94,gardiner2000}, a
formalism which deals with systems driven by non-classical
types of light and which was applied in the theory of Markovian
feedback \cite{wiseman1994b}.

The non-Markovian feedback contained in the system discussed here
makes it difficult to solve the problem in an exact analytical way
since it is not possible to establish a closed set of equations
describing the dynamics of the atom as in the theory of
Markovian resonance fluorescence. Thus we are restricted to approximative
methods in the following sections.  
\subsection{\label{sec:secIIIa}Perturbation theory}
Using the results of Sec.~\ref{sec:sec2} we will, as a starting point,
examine the influence of the laser for low intensities within the
framework of a time dependent perturbation theory. Here, as well as in
the following sections, the effect of the laser is included in our
considerations with the help of the standard semi-classical model for
atom-laser interaction in rotating wave approximation, i.e.\ the new
Hamiltonian reads
\begin{equation}
H_L = H + V(t),
\end{equation}
with
\begin{equation}
V(t) = - \hbar\frac{\Omega_0}{2}
              (\e^{\ii\omega_Lt}\sigma_- + \e^{-\ii\omega_Lt}\sigma_+)
\end{equation}
and laser- and Rabi-frequency $\omega_L$ and $\Omega_0$, respectively.
Taking the ground state of the atom-field system as initial state we
get in first order perturbation theory (assuming a weak laser
intensity) the excited state amplitude in a rotating frame
\begin{eqnarray}
b_e^L(t)&=& \e^{\ii\omega_Lt}\frac{1}{\ii\hbar}\int_0^t\dd t'\,
 \langle E\vert \e^{-\ii H(t-t')/\hbar}V(t')\e^{\ii Ht/\hbar}\vert G\rangle \nonumber\\
&=&\ii\frac{\Omega_0}{2}\int_0^t\dd t'\,\e^{-\ii\Delta t'}b_e(t') 
\label{eq:perturb}, 
\end{eqnarray}
with $\vert E\rangle\equiv\vert e,\{0\}_1,\{0\}_2\rangle$, $\vert
G\rangle \equiv \vert g,\{0\}_1,\{0\}_2\rangle$ and laser detuning
$\Delta\equiv\omega_0 - \omega_L$.  The above expression is
essentially equivalent to the one-photon amplitude of
Sec.~\ref{sec:sec2a} if the laser frequency $\omega_L$ is replaced by
$\omega$.  Thus, we immediately get
\begin{eqnarray}
b_e^L(t) &=& \frac{\ii\Omega_0}{\gamma +2\ii\Delta}
              \sum_{n=0}^{\infty}\frac{(\varepsilon\gamma/2)^n}{n!}
              \e^{\ii\omega_Ln\tau}(t-n\tau)^n \nonumber\\
&&\quad  \times G_n[-({\textstyle \frac{\gamma}{2}} + \ii\Delta)(t-n\tau)]
               \Theta(t-n\tau). \label{eq:perturb_result}
\end{eqnarray}
\begin{figure}[t]
  \centering \includegraphics[]{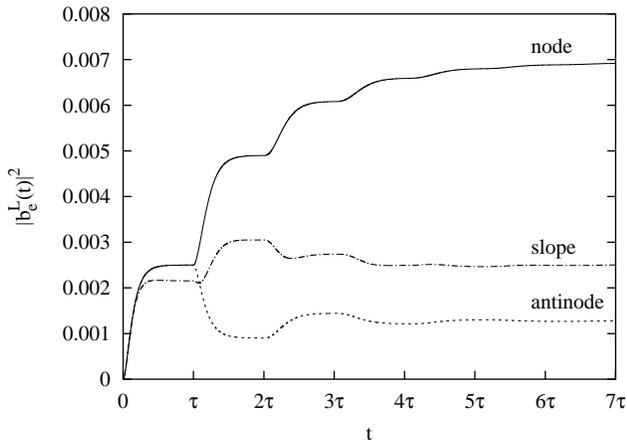}
 \caption{
   Upper state population in case of an atom far away from the mirror
   ($\gamma\tau = 20$, $\varepsilon=0.4$) and weak laser excitation
   ($\Omega_0 = 0.05\gamma$) for an atom placed in an antinode (dashed
   line, $\omega_L\tau=(2n+1)\pi$, $\Delta = 0$), in a node (solid
   line, $\omega_L\tau=2n\pi$, $\Delta = 0$) and at a ``slope''
   (dashed-dotted line, $\omega_L\tau=(2n-\frac{1}{2})\pi$, $\Delta
   =-0.2\gamma$).
\label{fig8}}
\end{figure}

Examples for the excited state amplitude are shown in Fig.~\ref{fig8}
for different positions of the atom whereas the overall distance of
the atom and the mirror is chosen to be quite large. The form of the
curves has a direct interpretation. Since for low laser intensities
coherent light scattering dominates the reflected radiation leads to a
lower or higher ``driving force'' depending on the position of the
atom. The system has similarities to an atom which is driven by two
lasers where the phase difference is controlled by the distance
between atom and mirror.  Actually the superimposed intensity of the
lasers changes after each round trip of the light which leads to a
transient upper state population as shown in Fig.~\ref{fig8}.  For
example, if the atom is placed in a node the laser interferes always
constructively with the ``reflected'' laser beam giving rise to a
higher population in a time interval $[n\tau,(n+1)\tau]$ compared to
the preceding one. This point will be further developed in the
Sec.~(\ref{sec:low_laser}) where we reconsider the limit of low laser
intensities.

The steady state population obtained from Eq.~(\ref{eq:perturb}) is
given by
\begin{equation}
\label{eq:spec_laser1}
\underset{t\rightarrow\infty}{\lim}\vert b_e^L(t) \vert^2 =
\frac{\Omega_0^2}{{\tilde\gamma_L}^2 + 4{\tilde\Delta}^2} 
\end{equation}
with modified decay rate and detuning
\begin{subequations}
\label{eq:mod_param}
\begin{eqnarray}
\tilde\gamma_L &=& \gamma(1-\varepsilon\cos(\omega_L\tau), \\
\tilde\Delta &=& \Delta - \varepsilon\frac{\gamma}{2}\sin(\omega_L\tau).
\label{eq:detuning_shift}
\end{eqnarray}
\end{subequations}
The situation is similar to the Markovian limit of the previous
section, i.e.\ we have a pronounced dependence of the atomic dynamics on the
exact position of the atom (e.g. node or antinode of a standing wave of
the laser frequency $\sin(k_Lz)$). The difference is that
this fact still holds in case of large atom-mirror
distances.  In the sense of Figs.~(\ref{fig3}) and (\ref{fig5})
this is due to the fact that the interference ability of outgoing and
reflected light does not depend on the distance since the
laser provides a continuous scattered light field.

A further quantity which is of interest in this context is the the
second order intensity correlation function,
\begin{equation}
\langle E^\dagger_j(t)E^\dagger_j(t+T)E_j(t+T)E_j(t)\rangle
= \vert\alpha_j\vert^4G^{(2)}_j(t,t+T),
\label{eq:G2}
\end{equation}
where the index $j=1,2$ indicates which channel is considered and
$\alpha_j\equiv(\delta_{j2}-\varepsilon)\gamma\hbar/(2d)$.
Expressions for the electric field operators are given in
Appendix~\ref{app:1}.  This correlation function corresponds to the
probability of detecting a photon at time $t+T$ on condition that at
time $t$ a first one was detected (see e.g.  \cite{glauber}).

Assuming that the detectors are arranged as in Fig.~\ref{fig11} we
have in the channel parallel to the mirror
\begin{eqnarray}
G^{(2)}_2(t,t+T) &=& \langle \sigma_+(t)\sigma_+(t+T)\sigma_-(t+T)\sigma_-(t)\rangle \nonumber\\
 &=& \| \sigma_-U(t+T,t)\sigma_-U(t,0)\vert G \rangle\|^2.
\end{eqnarray}
By calculating the time evolution $U$ in first order perturbation theory we
obtain
\begin{equation}
G^{(2)}_2(t,t+T) = \vert b_e^L(t)\vert^2\vert b_e^L(T)\vert^2,
\label{eq:G2t_perturb}
\end{equation}
or in the long time limit
\begin{equation}
\underset{t\rightarrow\infty}{\lim}G^{(2)}_2(t,t+T) 
= \frac{\Omega_0^2}{{\tilde\gamma_L}^2 + 4{\tilde\Delta}^2}\vert b_e^L(T)\vert^2,
\label{eq:G2_perturb}
\end{equation}
i.e.\ for $\gamma\tau\gg1$ we have a behavior as shown in
Fig.~\ref{fig8}. This can be interpreted as follows.  After the
detection of the first photon the atom is in its ground state and has
to be re-excited again before it is able to emit a second photon
(antibunching).  The radiation which is emitted now (the second
photon) is split up into a part emitted into channel $2$ and a part
which is emitted into channel $1$.  If $T<\tau$ the light which is (or
will be) reflected in channel $1$ is not able to reach the atom before
the photon detection in channel $2$.  Thus, we encounter, except for a
constant factor, the same behavior as in free space.  However, if
there is enough time for the radiation in channel $1$ to make a
complete round trip, it is able to re-interact with the atom (in
addition to the laser), which leads to a higher or lower emission
probability in channel $2$.
\begin{figure*}[t]
  \centering \includegraphics[]{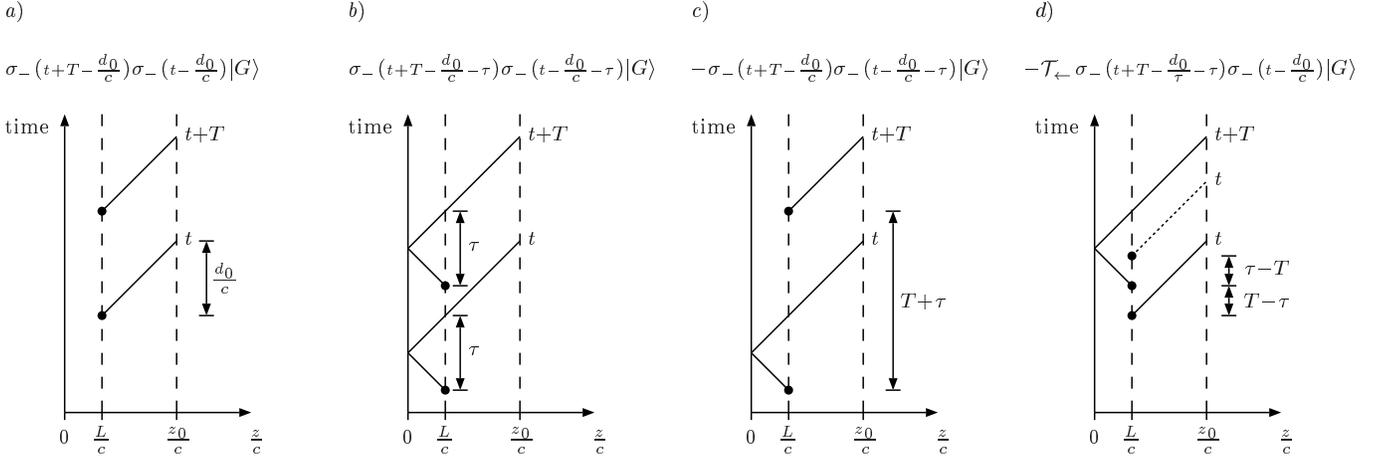}
 \caption{
   The four contributions to the second order intensity correlation
   function and the corresponding space-time diagrams. The correlation
   function is the squared norm of the sum of these vectors. The atom
   is located at $L$, the detector at $z_0$ and the distance between
   them is denoted as $d_0$. It is assumed that the detection of the
   first and second atom takes place at a time $t$ and $t+T$,
   respectively. The $\bullet$-symbol indicates the time of emission
   in the past relative to the detection times. The ordering of the
   operators in d) changes if $T$ becomes smaller than $\tau$ (dotted
   line).
\label{fig9}}
\end{figure*}

An expression for $G^{(2)}_1(t,t+T)$ is derived in
Appendix~\ref{app:1} where we assumed again a detector arrangement
as in Fig.~\ref{fig11}, i.e.\ the atom is located \textit{in between} the
mirror and the detector.  It is written as the norm of a sum of four
states.  These four contributions (in a non-rotating frame) are shown
in Figs.~\ref{fig9}a)-\ref{fig9}d) where each of
these terms is connected to a different path leading to a coincidence
detection at time $t$ and $t+T$. The back action of the light on the
atom is included in the dynamics of the time dependent operators.
In principle the light has two possibilities to get to the detector:
Either it takes the direct way or the indirect way via the mirror
which leads to four possibilities for a two-photon detection amplitude
indicated by the space-time diagrams in Fig.~\ref{fig9}.
For the sake of clarity the arbitrary distance between detector and
atom, $d_0$, is not set to zero which is also indicated in the time
arguments of the operators. For calculations however, we will always
set $d_0=0$.  The time arguments of the operators coincide with the
emission time in the past relative to $t$ and $t+T$, respectively.
Note, that a possibility which includes a reflection gives rise to a
negative sign and that the ordering of the operators in
Fig.~\ref{fig9}d) depends on the length of the delay
interval $T$.  The four contributions will interfere since they remain
indistinguishable when a coincidence signal occurs.

For weak laser intensities these quantities can be calculated using
first order perturbation theory in a similar way as it was done in the
derivation of Eq.~(\ref{eq:G2t_perturb}) which leads to an expression
of fourth order in $\Omega_0$,
\begin{align}
G_1^{(2)}(t,t+T) = \vert 
&b_e^L(T)b_e^L(t)+\e^{2\ii\omega_L\tau}b_e^L(T)b_e^L(t-\tau) \nonumber\\
&-\e^{\ii\omega_L\tau}b_e^L(T+\tau)b_e^L(t-\tau) \nonumber\\
&-\e^{\ii\omega_L\tau}b_e^L(\vert T-\tau\vert)b_e^L(t+s)\vert^2,
\label{eq:G_1(t)}
\end{align} 
where $s=0$ if $T>\tau$ and $s=T-\tau$ if
$T\le\tau$. We omitted step functions in this expression. In case of
negative arguments the corresponding quantities have to be set to
zero. Recall that these amplitudes are written in a rotating frame. In
the long time limit we have
\begin{align}
  &\underset{t\rightarrow\infty}{\lim}G_1^{(2)}(t,t+T)
  = \frac{\Omega_0^2}{{\tilde\gamma_L}^2 + 4{\tilde\Delta}^2} \nonumber\\
  &\qquad\times\vert 2b_e^L(T)\cos(\omega_L\tau) - b_e^L(T+\tau) -
  b_e^L(\vert T-\tau \vert) \vert^2.
 \label{eq:g_2^1}
\end{align}
\begin{figure}[t]
  \centering \includegraphics[]{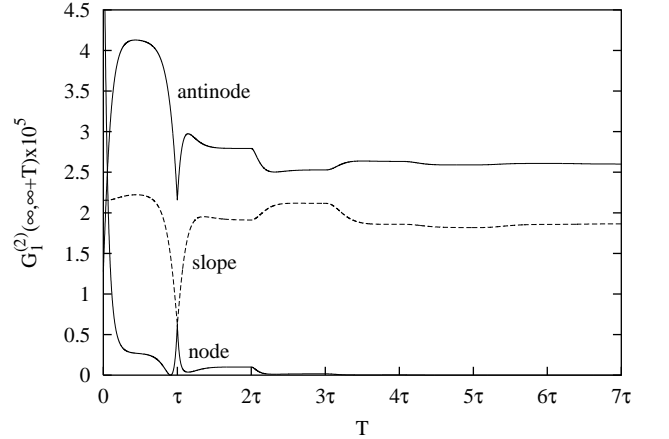}
 \caption{
   Second order intensity correlation function for a weakly driven
   atom ($\Omega_0 = 0.05\gamma$, $\varepsilon=0.4$) located at an
   antinode ($\omega_L\tau=(2n+1)\pi$), a ``slope''
   ($\omega_L\tau=(2n-\frac{1}{2})\pi$) and a node
   ($\omega_L\tau=2n\pi$) of the standing wave $\sin(k_Lz)$. The
   overall distance between the atom and the mirror is assumed to be
   quite large ($\gamma\tau = 20$) while the laser is tuned to exact
   resonance ($\Delta = 0$).
\label{fig10}}
\end{figure}
This function is shown in Fig.~\ref{fig10} for a relatively large
atom-mirror distance.  Several features are visible: First of all
$G_1^{(2)}$ is not zero for $T=0$.  Indeed, diagram a) and diagram b)
do not contribute to the detection probability in this case which
reflects the fact that after an emission process the atom is in its
ground state and the probability amplitude that it immediately emits a second
photon is zero.  On the other hand, if the first photon is detected
(and the atom is in the lower state) there is still the possibility
that there is radiation around caused by a prior emission process
which is represented by the remaining diagrams.  Due to this fact the
value of $G^{(2)}_1$ for $T=0$ differs from those for larger times
where the ``partial'' antibunching effect of a) and b) decreases. For
$T=\tau$ we have a similar situation concerning diagram d) which does
not contribute, i.e.\ a partial antibunching effect which leads again
to a different value of the detection probability compared to earlier
or later times.

Within the framework of this perturbative treatment we can also
calculate emission spectra without any effort which turn out to be
monochromatic. However, since the results coincide with those of
Sec.~\ref{sec:low_laser} they are not quoted here.

\subsection{Modified optical Bloch equations\label{sec:mod_bloch}}
For further investigations it turns out that it is advantageous to
work in the Heisenberg picture.

As the amplitudes in the previous sections, in the following the
atomic operators and the mode operators are always represented in a
rotating frame, i.e.\  $\sigma_- \rightarrow \e^{-\ii\omega_L
  t}\sigma_-$ and $a_k,\,b_k \rightarrow \e^{-\ii\omega_k
  t}a_k,\,\e^{-\ii\omega_k t}b_k$. The Heisenberg equations of motion
for the operators $a_k(t)$,
\begin{eqnarray}
\dot a_k(t) &=& g_k\sin(kL)\sigma_-(t)\e^{\ii(\omega_k - \omega_L)t},\label{eq:heis1}\\
\dot b_k(t) &=& h_k\sigma_-(t)\e^{\ii(\omega_k - \omega_L)t}
\end{eqnarray}
yield after formally integrating and inserting into
(\ref{eq:field_op}) and using a similar derivation as in
(\ref{eq:ww_approx}) the electric field operator at the position of
the atom,
\begin{align}
 E(t) =& \frac{\gamma}{2}\frac{\ii\hbar}{d}\e^{-\ii\omega_L t}\big( \sigma_-(t) 
  - \varepsilon  \e^{\ii\omega_L\tau}   \sigma_-(t-\tau)
  \Theta(t-\tau) \big) \nonumber\\
  & + N_1(t) + N_2(t)
\label{eq:e-field}
\end{align}
with $E(t)\equiv E_1(L,t) + E_2(0,t)$ and noise operators
\begin{eqnarray}
N_1(t) &=& \frac{\ii\hbar}{d}\int\dd k\,g_k\sin(kL)a_k(0)
                                      \e^{-\ii\omega_k t},\nonumber\\
N_2(t) &=& \frac{\ii\hbar}{d}\int\dd k\,h_k b_k(0)
                                      \e^{-\ii\omega_k t}.
\label{eq:noisy}
\end{eqnarray}
The second term on the right hand side of Eq.~(\ref{eq:e-field}),
\begin{equation}
E_{\text{ref}}(t) \equiv -\ii\varepsilon\frac{\gamma}{2}\frac{\hbar}{d}
                 \sigma_-(t-\tau)\e^{-\ii\omega_L(t-\tau)}\Theta(t-\tau) 
\label{eq:e-field_ref}
\end{equation}
can be identified as the reflected part of the electric
(source-)field.  Furthermore we will need in some calculations the
commutation relations of the noise operators and the ``atomic''
operators for $t'\le t$,
\begin{align}
&[N_j(t),\sigma_-(t')] =
  \varepsilon\frac{\gamma}{2}\frac{\ii\hbar}{d}
             \e^{-\ii\omega_L(t-\tau)}\delta_{j1} \nonumber\\
&\qquad\times[\sigma_-(t-\tau),\sigma_-(t')]\Theta(t-\tau)\Theta(t'-t+\tau).
\label{eq:commutator}
\end{align}
Note, that for $j=1$ this commutator is non-vanishing for $t'\le t\le
t'+\tau$ (and $t\ge\tau$) in contrast to the Markovian case.  With the
help of expression~(\ref{eq:e-field}), assuming again that all field
modes are initially in the vacuum state and keeping normal ordering of
the photon creation and annihilation operators, it is straightforward
to derive a set of modified optical Bloch equations (OBEs),
\begin{widetext}
\begin{eqnarray}
&&\frac{\dd}{\dd t}\langle\sigma_-\rangle = 
-\left(\frac{\gamma}{2} + \ii\Delta\right)\langle\sigma_-\rangle -\ii\frac{\Omega_0}{2}
\big( \langle\sigma_+\sigma_-\rangle - \langle\sigma_-\sigma_+\rangle \big) 
 -\varepsilon\frac{\gamma}{2}\e^{\ii\omega_L\tau}
      \big( 
           \langle\sigma_+\sigma_-\sigma_-(t-\tau)\rangle 
               - \langle\sigma_-\sigma_+\sigma_-(t-\tau)\rangle
       \big) \Theta(t-\tau), \nonumber\\
&&\frac{\dd}{\dd t}\langle{\sigma_+\sigma_-}\rangle = 
\ii\frac{\Omega_0}{2} \big( \langle\sigma_+\rangle - \langle\sigma_-\rangle \big) 
- \gamma\langle\sigma_+\sigma_-\rangle + \varepsilon\frac{\gamma}{2} 
   \big( \e^{-\ii\omega_L\tau}\langle\sigma_+(t-\tau)\sigma_-\rangle 
           + \e^{\ii\omega_L\tau}\langle\sigma_+\sigma_-(t-\tau)\rangle   
   \big)\Theta(t-\tau), \nonumber \\
&&\frac{\dd}{\dd t}\langle\sigma_+\rangle 
         = \left( \frac{\dd}{\dd t}\langle\sigma_-\rangle \right)^*, 
\quad\frac{\dd}{\dd t}\langle{\sigma_-\sigma_+}\rangle
         = -\frac{\dd}{\dd t}\langle\sigma_+\sigma_-\rangle, \label{eq:dgl2}
\end{eqnarray}
\end{widetext}
where we indicate for the sake of clarity only the retarded time
arguments.  As can be seen from these equations, the non-linear
structure of the Heisenberg equations of motion leads to the
appearance of correlation functions on the right hand side of
Eq.~(\ref{eq:dgl2}). Hence, the modified OBEs~(\ref{eq:dgl2}) can not
be considered as a closed set of equations. However, it is convenient
to take them as a starting point for approximative treatments.

\subsubsection{Small distance between atom and mirror (Markov limit)\label{sec:close}}
As in Sec.~\ref{sec:sec2a} we will first consider the limit of small
distances between the atom and the mirror, i.e.\  $\gamma\tau\ll1$.
Furthermore, we require now that the intensity and the
detuning of the laser is not too high which can be expressed by the
condition $\Omega\tau\ll1$ with
$\Omega\equiv\sqrt{\Omega_0^2+\Delta^2}$.  The latter defines a time
scale on which the solution of the usual OBEs ($\varepsilon=0$) vary
appreciably in the high intensity limit \cite{eberly}, 
and we suppose that it will give us also an estimation of this scale
in case of $\varepsilon\ne0$. In Sec.~\ref{sec:high_laser} it will be
shown that this condition has also some meaning in a frequency space.
Thus, we can make again the approximation $\tau\rightarrow +0$ in the
arguments of the operators in Eq.~(\ref{eq:dgl2}) which leads to the
equations
\begin{eqnarray}
\frac{\dd}{\dd t}\langle\sigma_-\rangle 
&=& -\left( \frac{\tilde\gamma_L}{2} + \ii\tilde\Delta\right) 
    \langle\sigma_-\rangle -\ii\frac{\Omega_0}{2}\langle\sigma_z\rangle, \nonumber\\
\frac{\dd}{\dd t}\langle\sigma_+\rangle 
&=& -\left( \frac{\tilde\gamma_L}{2} - \ii\tilde\Delta \right)
    \langle\sigma_+\rangle + \ii\frac{\Omega_0}{2}\langle\sigma_z\rangle, \nonumber\\
\frac{\dd}{\dd t}\langle\sigma_z\rangle 
&=& \ii\Omega_0\big(\langle\sigma_+\rangle - \langle\sigma_-\rangle\big) 
    - \tilde\gamma_L\big(\langle\sigma_z\rangle + 1\big), \label{eq:bloch_approx}
\end{eqnarray}
where $\tilde\gamma_L$ and $\tilde\Delta$ are defined by
Eq.~(\ref{eq:mod_param}). For the sake of simplicity we set $\tau=0$
also in the arguments of the step functions (in contrast to
Eq.~(\ref{eq:dgl1})), i.e.\ the difference in the dynamics in the time
interval $[0,\tau]$ and later is neglected.  Anyhow, taking the
difference into account would merely lead to a (slightly) different
initial condition for Eq.~(\ref{eq:bloch_approx}) which would not
alter the steady state results to be discussed here at all.

Thus, the equations have the form of the familiar OBEs with modified
spontaneous emission rate and detuning. They are rewritten as an
inhomogeneous system of three differential equations where $\sigma_z
\equiv \sigma_+\sigma_- - \sigma_-\sigma_+$, since it is more
convenient to perform steady state calculations in this
representation.

The steady state population of the upper state is easily obtained by
inverting a $3\times3$-matrix which corresponds to
Eq.~(\ref{eq:bloch_approx}),
\begin{equation}
\label{eq:steady1}
\langle\sigma_+\sigma_-\rangle_{ss} 
= \frac{\Omega^2_0}{ {\tilde\gamma_L}^2 + 2\Omega_0^2 + 4{\tilde\Delta}^2}.
\end{equation}
Understood as a function of $\Delta$ this is essentially a Lorentzian
(in the limit under consideration, treating the trigonometric
functions as constants) with maximum at $\Delta_{\text{max}} =
\varepsilon\frac{\gamma}{2}\sin(\omega_L\tau)$ and width $w =
\sqrt{{\tilde\gamma_L}^2 + 2\Omega_0^2}$.  Applying again the standing
wave picture of Sec.~\ref{sec:sec2a} we see that the shift of the
maximum vanishes if the atom is located at a node or an antinode of
$\sin(\omega_L\tau/2)$. In contrast to this, the width takes its
minimum or maximum values at these points. Indeed we get a maximum
shift if the atom is placed exactly \textit{in between} a node and an
antinode where the spontaneous emission rate is not altered at all.
\begin{figure}[t]
  \centering \includegraphics[]{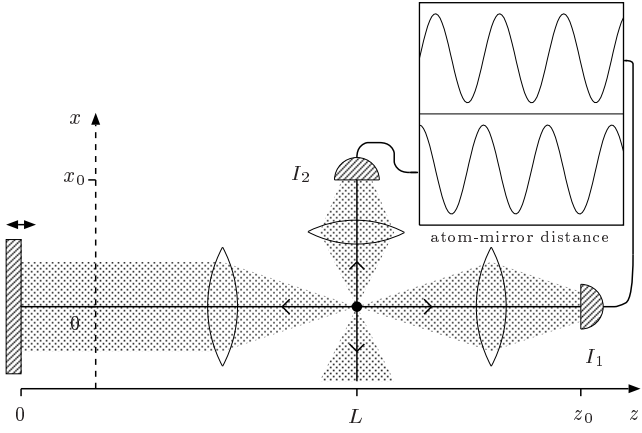}
 \caption{
   Sketch of the detector arrangement. A first detector opposite to
   the mirror ($z_0>L$) measures the intensity $I_1$ in the first
   channel while a second detector measures the intensity $I_2$ of the
   light in the channel parallel to the mirror.
\label{fig11}}
\end{figure}
In the limit discussed here the steady state population is
proportional to the measurable intensities
\begin{subequations}
\label{eq:int12}
\begin{eqnarray}
I_1&\sim& \sin^2(\omega_L\tau/2)\langle\sigma_+\sigma_-\rangle_{ss},
 \label{eq:intensity0}\\
I_2&\sim& \langle\sigma_+\sigma_-\rangle_{ss} 
\end{eqnarray}
\end{subequations}
of the light emitted in channel $1$ or $2$, respectively (see
Fig.~\ref{fig11} and Appendix~\ref{app:1}). Thus, a possible way to
demonstrate effects caused by the mirror is to measure the
absorption spectrum of the atom, i.e.\ the intensity of the scattered
light depending on the laser detuning.  A further option would be the
measurement of the intensity $I_2$ for different positions of the
mirror (keeping $\Delta$ constant), i.e.\  for different values of
$\tau$, which would lead to a periodic variation in the measured
intensity. This was done (for small global atom-mirror distances) in
\cite{eschner} whereas the measurement scheme slightly differed from
that discussed here since the system under consideration was a
three-level atom \footnote{In the experiment \cite{eschner} the
  $\Lambda$-type three-level system was excited by two lasers where
  the mirror affected merely the radiation of one transition. Thus, we
  have essentially a completely (``free space''-)Markovian behavior
  concerning the radiation originated by the other transition. The
  intensities of the light at both of these frequencies was measured
  simultaneously and it is clear that, like the light in channel $2$,
  the intensity of the non-reflected light is also simply proportional
  to the upper state population. For small $\varepsilon$ we have
  essentially the same oscillatory behavior whereas the expressions
  for the amplitude and the phase shift of the oscillations are much
  more complicated.}.  However, the basic principle is the same and
for effects discussed in this paper it is sufficient to consider a
two-level system.

If we assume that $\varepsilon\ll1$, Eq.~(\ref{eq:steady1}) can be
expanded to lowest order in this parameter,
\begin{equation}
\langle\sigma_+\sigma_-\rangle_{ss} 
\approx \frac{\Omega_0^2}{\Gamma}\left( 1 + 2\varepsilon\frac{\gamma^2}{\Gamma}
\sqrt{\frac{\gamma^2+4\Delta^2}{\gamma^2}}\cos(\omega_L\tau-\varphi)  \right) 
\label{eq:state_oscillation_near}
\end{equation}
with $\Gamma\equiv \gamma^2+2\Omega_0^2+4\Delta^2$ and $\tan(\varphi) =
2\Delta/\gamma$.  The presence of the relatively small level
shift leads to a $\Delta$-dependent phase shift $\varphi$ with respect
to the function $\cos(\omega_L\tau)$ which corresponds to the phase of
the standing wave $\sin(\omega_L\tau/2)$. The determination of this
phase would, e.g., require the knowledge of the exact distance between
atom and mirror. This difficulty could be avoided if one carries out a
simultaneous measurement of $I_2$ and $I_1$ since the phase of the
latter is dominated for small $\varepsilon$ by the prefactor
$\sin^2(\omega_L\tau/2)=(1-\cos(\omega_L\tau))/2$. This means that, if
there would be no level shift or $\Delta=0$, the two signals were
anticorrelated (i.e.\ a minimum of the $I_1$-signal coincides with
a maximum of the $I_2$-signal). The existence of a level shift
removes, in case of a finite detuning, this coincidence (see inset in
Fig.~\ref{fig11}).
\begin{figure}[t]
  \centering \includegraphics[]{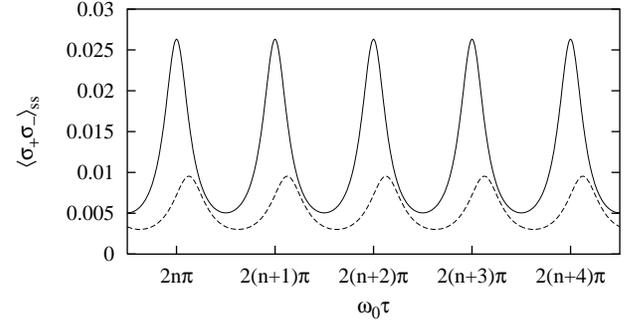}
 \caption{
   Excited state population depending on the distance between atom and
   mirror for $\Delta=0$ (solid line) and $\Delta=\gamma/2$ (dashed
   line) and a larger solid angle ($\varepsilon=0.4$). The global
   distance is assumed to be small while the laser intensity is weak
   ($\Omega_0=0.1\gamma$).
\label{fig12}}
\end{figure}
For higher values of $\varepsilon$ we can get deviations from a
pure sinusoidal behavior (see Fig.~\ref{fig12}).

\subsubsection{Low laser intensity\label{sec:low_laser}}
Provided with Eq.~(\ref{eq:dgl2}) we can now also examine the limit of
small laser intensities in more detail. In this case (assuming that
the atom is initially in the ground state) one expects that the
atomic operators are approximately uncorrelated since coherent
scattering processes dominate, i.e.\ we can make substitutions of the
type
\begin{equation}
\langle\sigma_q(t)\sigma_{q'}(t')\rangle \approx \langle
       \sigma_q(t)\rangle\langle\sigma_{q'}(t')\rangle. 
\label{eq:assump1}
\end{equation}
After some rearrangements one gets
\begin{align}
  \frac{\dd}{\dd t}\langle\sigma_-\rangle &= -\left(\frac{\gamma}{2} +
    \ii\Delta\right)\langle\sigma_-\rangle
  -\ii\frac{1}{2}\Pi(t)\langle\sigma_z\rangle, \nonumber\\
  \frac{\dd}{\dd t}\langle\sigma_+\sigma_-\rangle &=
  \frac{\ii}{2}\big(\Pi(t)\langle\sigma_+\rangle -
  \Pi^*(t)\langle\sigma_-\rangle\big) -
  \gamma\big(\langle\sigma_+\sigma_-\rangle\big),
\label{eq:dgl_lowlaser}
\end{align}
where we introduced the quantity
\begin{equation}
\Pi(t) = \Omega_0 - \ii\varepsilon\gamma\e^{\ii\omega_L\tau}
         \langle \sigma_-(t-\tau)\rangle\Theta(t-\tau). \label{eq:Pi1}
\end{equation}
With the help of the decorrelation assumption (\ref{eq:assump1}) we eliminate the field
degrees of freedom which are implicitly still contained in
Eq.~(\ref{eq:dgl2}) and get an equation for a reduced atomic system.
This assumption is related to the fact that an atom initially in the
ground state and weakly excited by a laser approximately behaves like
a harmonic oscillator since $\sigma_z=[\sigma_+,\sigma_-]\approx
-\openone$.  With regard to Eq.~(\ref{eq:dgl_lowlaser}) we have
to replace $\langle\sigma_z\rangle$ by $-1$ and it can be shown (see
Appendix~\ref{app:2}) that assumption (\ref{eq:assump1}) holds in this
case if the system is initially in the ground state.

Anyhow, for the following discussion we keep for a short time the
$\sigma_z$-term since the equations are more transparent in this form
because the principal form of OBEs is conserved.  Apparently the
quantity $\Pi(t)$ can be interpreted as a modified Rabi-frequency
in particular if we recall the form of that part of the electric field
operator which is due to the reflection of the light (see
Eq.~(\ref{eq:e-field_ref})), $\langle
E_{\text{ref}}(t)\rangle\approx E_0(t)\e^{-\ii\omega_L t}$, with a
slowly varying amplitude $E_0(t)$.  Thus, Eq.~(\ref{eq:Pi1}) can be
written in the form $\Pi(t) = \Omega_0 + (2d/\hbar)E_0(t)$, where the
last term coincides with the definition of a Rabi-frequency.  Let us
assume now that $\gamma\tau\gg1$ and that the atom is in the ground
state at $t=0$.  Then, the modified Rabi-frequency (\ref{eq:Pi1}) has
approximately the shape of a ``stair function'' (going up and down in
general) with mostly decreasing distance between the single steps.
This can be understood if we discuss the time evolution of the system
in time intervals of length $\tau$: Between $t=0$ and $t=\tau$
Eq.~(\ref{eq:dgl_lowlaser}) are the ordinary OBEs with Rabi-frequency
$\Omega_0$ since the Heaviside function in (\ref{eq:Pi1}) vanishes.
The solution of the ordinary OBEs yields for $\Omega_0\ll\gamma$
\begin{equation}
\langle\sigma_-(t)\rangle \approx \frac{\ii\Omega_0}{\gamma + 2\ii\Delta},
                  \quad t\in[0,\tau],
\end{equation} 
because the above expectation value is effectively constant after a
few radiative lifetimes $1/\gamma\ll\tau$.  According to this, in the
next time interval $[\tau,2\tau]$, the Rabi-frequency takes the form
\begin{equation}
\Omega_0' = \Omega_0(1+\mu\e^{\ii\omega_L\tau}) \quad \text{with} 
            \quad \mu\equiv \frac{\varepsilon\gamma}{\gamma + 2\ii\Delta}.
\end{equation}
Now we have to solve again the ordinary OBEs which leads to
\begin{equation}
\langle\sigma_-(t)\rangle \approx \frac{\ii\Omega_0'}{\gamma + 2\ii\Delta},
                  \quad t\in[\tau,2\tau],
\end{equation} 
giving rise to a new Rabi-frequency $\Omega_0''$ in $[2\tau,3\tau]$
and so on. Thus, Eq.~(\ref{eq:dgl_lowlaser}) takes in every time
interval $[n\tau,(n+1)\tau],\,n\in\mathbb{N}_0$ the form of ordinary
OBEs with different Rabi-frequencies $\Omega_0^{(n)}$ defined by
\begin{equation}
\Omega_0^{(n)} = \Omega_0 + \mu\e^{\ii\omega_L\tau}\Omega_0^{(n-1)},
                 \quad\Omega_0^{(0)} = \Omega_0,
\label{eq:recurrence}
\end{equation}
while the expectation value of the dipole operator in an
``intermediate'' steady (cf. Fig.~\ref{fig8}) state is given by
\begin{equation}
\langle\sigma_-\rangle^{(n)}
       \approx\frac{\Omega_0^{(n)}}{\gamma + 2\ii\Delta},
       \quad t\in[n\tau,(n+1)\tau].
\label{eq:inter}
\end{equation}
The value for $n\rightarrow\infty$ of the recurrence
relation~(\ref{eq:recurrence}) is the limit of a geometric series or
simply the fixed point of the map which is given by
\begin{equation}
\Omega_0^{(\infty)} = \frac{\Omega_0}{1-\mu\e^{\ii\omega_L\tau}},
\end{equation}
and leads to a steady state population
$\langle\sigma_+\sigma_-\rangle_{ss}\approx\vert\Omega_0^{(\infty)}\vert^2/(\gamma^2
+ 4\Delta^2)$ which coincides exactly with expression
(\ref{eq:spec_laser1}) we got from the perturbation theory.

Apart from this semi quantitative discussion it is worthwhile to study
the equations of motion~(\ref{eq:dgl_lowlaser}) in more detail.  We
explicitly make now the replacements $\sigma_-\rightarrow c$,
$\sigma_+\rightarrow c^\dagger$ and therefore $\sigma_z\rightarrow -\openone$,
where $c$ is a lowering operator of a harmonic oscillator.  The
equations of motion then take the form
\begin{subequations}
\begin{eqnarray}
\frac{\dd}{\dd t}\langle c(t) \rangle &=& 
-\left( \frac{\gamma}{2}+ \ii\Delta \right)\langle c(t)\rangle 
                        + \ii\frac{\Omega_0}{2} \nonumber\\
&&\qquad+ \varepsilon\frac{\gamma}{2}\e^{\ii\omega_L\tau}
          \langle c(t-\tau)\rangle\Theta(t-\tau),
\label{eq:DDGL_harm}
\\
\frac{\dd}{\dd t}\langle c^\dagger(t) c(t) \rangle 
&=& -\ii\frac{\Omega_0}{2} 
      \big( \langle c(t)\rangle - \langle c^\dagger(t)\rangle \big) 
    -\gamma\langle c^\dagger(t) c(t)\rangle  \nonumber\\
&&\quad + \varepsilon\frac{\gamma}{2}
   \big( \e^{-\ii\omega_L\tau}\langle c^\dagger(t-\tau)c(t)\rangle \nonumber\\
&&\quad\qquad +\e^{\ii\omega_L\tau}\langle c^\dagger(t)c(t-\tau)\rangle     
                   \big)\Theta(t-\tau). \nonumber\\
\label{eq:DDGL_harmb}
\end{eqnarray}
\end{subequations}
In Appendix~\ref{app:2} it is shown that both $\langle c^\dagger
(t)c(t)\rangle=\langle c^\dagger(t)\rangle\langle c(t)\rangle$ and, as
already mentioned, also the two-time correlation functions factorize
if the atom is initially in the ground state.  Furthermore, it is
proved that these statements still hold in a steady state regime
independently of the initial state.

Our task now is merely to solve Eq.~(\ref{eq:DDGL_harm}) assuming an
initially unexcited atom.  This equation is a linear delay
differential equations, i.e.\, apart from the constant inhomogeneity, a
type of equation like it already appeared in Sec.~\ref{sec:sec2}.  As
a matter of fact, this equation reduces in every time interval
$[n\tau,(n+1)\tau]$ to a ordinary linear differential equation (ODE)
with a time dependent inhomogeneity, so for any given initial state
there exists a unique solution. In contrast to initial value problems
concerned with ODE, delay differential equations need an initial
\textit{function}. In our case this initial function is defined due to
the presence of the step function which yields in a first time
interval $[0,\tau]$ an ODE and is of of course given by its solution.
This initial function is uniquely defined by the initial state and it
replaces the quantity $\langle c(t-\tau)\rangle$ in the equation of
motion in the next time interval $[\tau,2\tau]$ leading to an ODE with
a time dependent inhomogeneity. As initial value we take of course the
solution of the ODE in the first interval at $t=\tau$ (which is justified 
since it can be easily shown that the solutions of the type of
equations we consider here have to be continuous). The solution in
$[\tau,2\tau]$ provides us again with the functions $\langle
c(t-\tau)\rangle$ in $[2\tau,3\tau]$ and an initial value. Continuing
this procedure, we see that we have to solve in every time interval
$[n\tau,(n+1)\tau]$ an initial value problem of ODE and we can apply
all mathematical theorems which are concerned with such kind of
equations. This ``method of steps'' \cite{driver} can even yield
analytical solutions as we will see in Sec.~\ref{sec:high_laser}.
Another method in order solve linear delay differential equations is
by Laplace transformation like it was done in case of
Eq.~(\ref{eq:dgl1}) since in Laplace space the function with the
retarded time argument is simply replaced by the Laplace transformed
of that function multiplied by an exponential function.

Thus, the solution of Eq.~(\ref{eq:dgl_lowlaser}) is
unique (for a given initial state) whereas the behavior of the
derivatives is more complicated.  With respect to this, it can
be easily shown that (under certain conditions which are fulfilled in
our case) the solution has at least $n$th continuous derivatives at
$t=n\tau$ and in general the $(n+1)$th derivative has a discontinuity.
This feature can be identified in Fig~\ref{fig2}, for instance,
where we recognize a kink at $t=\tau$.

In order to demonstrate the mentioned solution method we will now
derive the transient solution of the delay differential
equation~(\ref{eq:DDGL_harm}). The Laplace transformed of the
expectation value $\langle c\rangle$ takes the form (assuming $\langle
c(0)\rangle=0$)
\begin{equation}
\mathfrak{L}[\langle c(t)\rangle](\ii z+\xi) = \frac{\alpha_3}{\ii z +\xi}
             \frac{1}{\ii z + \xi + \alpha_1 -\alpha_2\e^{-(\ii z+\xi)\tau}}
\end{equation}
with $\alpha_1\equiv-\left( \frac{\gamma}{2}+\ii\Delta \right)$,
$\alpha_2\equiv\varepsilon\frac{\gamma}{2}\e^{\ii\omega_L\tau}$,
$\alpha_3\equiv\ii\frac{\Omega_0}{2}$, and $\xi\in\mathbb{R}^+$. 
We get 
\begin{eqnarray}
\langle c(t)\rangle &=& 
\frac{\alpha_3}{2\pi}\int_{-\infty}^\infty \dd z\,\frac{\e^{(\ii z+\xi)t}}{\ii z +\xi}
\frac{1}{1-\frac{\alpha_2\e^{-(\ii z +\xi)\tau}}{\ii z + \xi + \alpha_1}} 
\frac{1}{\ii z + \xi + \alpha_1} \nonumber\\
&=& 
\frac{\alpha_3}{2\pi}\sum_{n=0}^\infty\alpha_2^n\e^{\ii\xi(t-n\tau)} \nonumber\\
&&\qquad\quad\times\int_{-\infty}^\infty\dd z\,
    \frac{\e^{\ii z(t-n\tau)}}{\ii z +\xi}\frac{1}{(\ii z +\xi+\alpha_1)^{n+1}}. 
\nonumber\\\label{eq:demo}
\end{eqnarray}
The Fourier transformation in the last line of this expression (see
e.g. \cite{magnus}) leads finally to the result
(\ref{eq:perturb_result}) already obtained from perturbation theory
and a closer inspection of it confirms the result
(\ref{eq:inter}).

The upper state population in a stationary regime coincides therefore
exactly with Eq.~(\ref{eq:spec_laser1}) which is the low intensity
limit of Eq.~(\ref{eq:steady1}). However, Eq.~(\ref{eq:spec_laser1})
is also valid for $\gamma\tau\gg1$.  The intensities in a stationary
regime measured in channel one and two, respectively (see
Fig.~\ref{fig11}) take again the form~(\ref{eq:int12}) which is due to
the factorization property of the two-time correlation functions.
Using this fact, we can furthermore easily calculate emission spectra
of the light scattered in channel $1$ or $2$ which gives
\begin{subequations}
\label{eq:harm_spectra}
\begin{eqnarray}
S_1(\omega)
&\sim& \sin^2(\omega_L\tau/2)\vert\langle c\rangle_{ss}\vert^2
                                        \delta(\omega-\omega_L),\\
S_2(\omega)
&\sim& \vert\langle c\rangle_{ss}\vert^2\delta(\omega-\omega_L).
\end{eqnarray}
\end{subequations}
The fact that the spectra are monochromatic just expresses again that
coherent, elastic scattering processes are involved in the
limit of low laser intensities.

It was already mentioned that the second order correlation functions
factorize under certain circumstances but what about higher order
correlation functions?  A lack of the harmonic oscillator model is
surely that in general the operator $c(t)^2$ is not equal to zero in
contrast to $\sigma_-(t)^2$ so we cannot necessarily expect that for
example the quantity
\begin{equation}
G^{(2)}_2(t,t+T) = \langle c^\dagger(t)c^\dagger(t+T)c(t+T)c(t)  \rangle
\end{equation}
gives the correct result for $T\rightarrow0$. In fact it can be seen
from the results of Appendix~\ref{app:2} that $G^{(2)}_j(t,t+T)$ is in
general not equal to zero for $T=0$, so as in the theory of ordinary
resonance fluorescence (see, e.g., \cite{knight80}) the correct result is
obtained by perturbation theory.

A further remarkable fact is that the harmonic oscillator model
reproduces the result of the Wigner-Weisskopf theory of
Sec.~\ref{sec:sec2} where pure spontaneous decay was considered (see
Appendix~\ref{app:2}). At a first glance this seems to be surprising
since the atom was initially in the excited state, i.e.\
$\langle\sigma_z\rangle$ was far away from $-1$.  On the other hand we
saw in Sec.~\ref{sec:model} that the state of the system is in this
case always confined to the subspace spanned by the vectors $\{\vert
e,\{0\}_1,\{0\}_2\rangle,\vert g,\{k\}_1,\{0\}_2\rangle,\vert
g,\{0\}_1,\{k\}_2\rangle\}$ (and $\vert g,\{0\}_1,\{0\}_2\rangle$ if
one wants to start in a state different from the excited state) which
leads to the fact that the noise terms of the Heisenberg equations
still do not contribute to the modified OBEs. Furthermore the two-time
correlation function takes the form $\langle
\sigma_+(t')\sigma_-(t)\rangle = b_e^*(t')b_e(t)$ and thus, the
equation of motion (from Eq.~(\ref{eq:dgl2})) for the upper state
probability is equal to Eq.~(\ref{eq:DDGL_harmb}) for vanishing laser
intensity.

\subsubsection{High laser intensity\label{sec:high_laser}}
The examination of the systems dynamics for larger values of
$\Omega_0\tau$ is more complicated since the incoherent nature of the
scattered (and reflected) radiation becomes important.  In order to
investigate the dynamics in this parameter regime we will assume in
the following that $\varepsilon$ is small so we can treat the
``reflected'' part of Eq.~(\ref{eq:dgl2}) as a perturbation. With the
aim to obtain a closed set of equations which contain only terms of
first order in $\varepsilon$ we can calculate the two-time correlation
functions in $0$-th order $\varepsilon$ depending on the initial state
which is a single time expectation value and reinsert the result
into~(\ref{eq:dgl2}). To this end we can multiply the Heisenberg
equations of motion with $\sigma_+(t')$ from the left or
$\sigma_-(t')$ from the right where $t'\le t$ , make use of the
commutation relations (\ref{eq:commutator}) and calculate the
expectation value. The equations we get in this way now contain third
order correlation functions which are, however, of order
$\varepsilon$, and thus they are neglected.  
The solution for $t'=t-\tau$ is given by
\begin{equation}
\vec C_\pm(t,t-\tau) = U(\tau) \vec C_\pm(t-\tau,t-\tau)
\end{equation}
with
\begin{eqnarray}
\vec C_+(t,t') &\equiv& 
      \begin{pmatrix} \langle\sigma_+(t')\sigma_-(t)\rangle       \\ 
        \langle\sigma_+(t')\sigma_+(t)\rangle        \\
        \langle\sigma_+(t')\sigma_+(t)\sigma_-(t)\rangle \\
        \langle\sigma_+(t')\sigma_-(t)\sigma_+(t)\rangle
      \end{pmatrix}, \\ 
\vec C_-(t,t') &\equiv& 
      \begin{pmatrix} \langle\sigma_-(t)\sigma_-(t')\rangle       \\ 
        \langle\sigma_+(t)\sigma_-(t')\rangle        \\
        \langle\sigma_+(t)\sigma_-(t)\sigma_-(t')\rangle \\
        \langle\sigma_-(t)\sigma_+(t)\sigma_-(t')\rangle
      \end{pmatrix}. 
\end{eqnarray}
The matrix elements $U_{ij}(\tau)$ of the evolution operator $U(\tau)
= \e^{A_4\tau}$ with
\begin{equation}
A_4 \equiv \begin{pmatrix} 
 -\frac{\gamma}{2}-\ii\Delta  & 0 & -\ii\frac{\Omega_0}{2} & \ii\frac{\Omega_0}{2} \\ 
 0 & -\frac{\gamma}{2}+\ii\Delta & \ii\frac{\Omega_0}{2} & -\ii\frac{\Omega_0}{2} \\
 -\ii\frac{\Omega_0}{2} & \ii\frac{\Omega_0}{2} & -\gamma & 0 \\
 \ii\frac{\Omega_0}{2} & -\ii\frac{\Omega_0}{2} & \gamma & 0 
\end{pmatrix}
\end{equation}
are obtained by solving the corresponding differential equation.  

By inserting the $0$-th order two-time correlation functions into
Eq.~(\ref{eq:dgl2}) we get finally an equation which is of first order
in $\varepsilon$,
\begin{equation}
\dot{\vec S}(t) = A_4\vec S(t) + \varepsilon\,K(\tau)\vec S(t-\tau)\Theta(t-\tau),
\label{eq:DDGL3}
\end{equation}
where we introduced the abbreviations
\begin{equation}
\vec S(t) \equiv 
\big(\langle\sigma_-(t)\rangle,       
     \langle\sigma_+(t)\rangle,        
     \langle\sigma_+(t)\sigma_-(t)\rangle, 
     \langle\sigma_-(t)\sigma_+(t)\rangle\big)^T
 ,
\end{equation}
\begin{equation}
K(\tau) \equiv 
\begin{pmatrix} 
  \frac{\gamma}{2}f_1(\tau) & 0 & -\ii\frac{\Omega_0}{2}f_2(\tau)
  & 0 \\
  0 & \frac{\gamma}{2}f_1^*(\tau) & \ii\frac{\Omega_0}{2}f_2^*(\tau)
  & 0 \\
  -\ii\frac{\Omega_0}{2}f_3(\tau) & \ii\frac{\Omega_0}{2}f_3^*(\tau) &
  \gamma f_4(\tau)
  & 0 \\
  \ii\frac{\Omega_0}{2}f_3(\tau) & -\ii\frac{\Omega_0}{2}f_3^*(\tau) &
  -\gamma f_4(\tau) & 0
\end{pmatrix}
\label{eq:matrix_K}
\end{equation}
with
\begin{eqnarray}
f_1(\tau) &=& -\e^{\ii\omega_L\tau}[U_{34}(\tau) - U_{44}(\tau)], \label{eq:element1}\\
f_2(\tau) &=& -\e^{\ii\omega_L\tau}\frac{2\ii\gamma}{\Omega_0}U_{31}^*(\tau),\\
f_3(\tau) &=& \e^{\ii\omega_L\tau}\frac{\ii\gamma}{\Omega_0} U_{24}(\tau),\label{eq:element3} \\
f_4(\tau) &=& \frac{1}{2}(\e^{-\ii\omega_L\tau}U_{11}(\tau) + \e^{\ii\omega_L\tau}U_{11}^*(\tau)).
\end{eqnarray}
Obviously, Eq.~(\ref{eq:DDGL3}) describes again a reduced atomic
dynamics but compared to the equation discussed in
Sec.~\ref{sec:low_laser} it is more complicated since we have now a
coupled system of four delay differential equations. These equations
are an extension of the ordinary OBEs (which are recovered in the
$\varepsilon\rightarrow 0$ limit).  We can apply the method of steps
which yields the formal solution for times $t\in[m\tau,(m+1)\tau]$,
\begin{eqnarray}
\vec S(t) &=& U(t)\vec S(0) + \sum_{n=1}^m (\varepsilon\gamma)^mU(t)\int_{m\tau}^t\dd t_1\, 
            \int_{(m-1)\tau}^{t_1-\tau}\dd t_2 \nonumber\\
&&\quad\ldots\int_{\tau}^{t_{m-1}-\tau}\dd t_m\,B(t_1)B(t_2)\ldots B(t_m)\vec S(0), 
\nonumber\\&&\label{eq:sol2}
\end{eqnarray}
where $B(t)\equiv U^{-1}(t)K(\tau)U(t-\tau)$. The above expression has
a form similar to that of the excited state amplitude~(\ref{eq:dgl1}).
In fact, Eq.~(\ref{eq:sol2}) yields in case of vanishing laser
intensity
\begin{eqnarray}
\langle\sigma_+(t)\sigma_-(t) \rangle 
&=& \sum_{n=0}^\infty \frac{(\varepsilon\gamma)^n}{n!}
               \cos^n(\omega_0\tau)(t-n\tau)^n \nonumber\\
&&\qquad\qquad \times \e^{-\gamma(t-n\frac{\tau}{2})}\Theta(t-n\tau)
\end{eqnarray}
which is a acceptable approximation for
$\varepsilon\gamma\tau\ll1$. Furthermore, if we assume that
$\gamma\tau\ll1$, $\Omega\tau\ll1$ so that $U(\tau)\approx\openone$
and $\tau\rightarrow +0$ in the arguments of $\vec S$, we recover
Eq.~(\ref{eq:bloch_approx}) of Sec.~\ref{sec:close}.

\begin{figure}[t]
  \centering \includegraphics[]{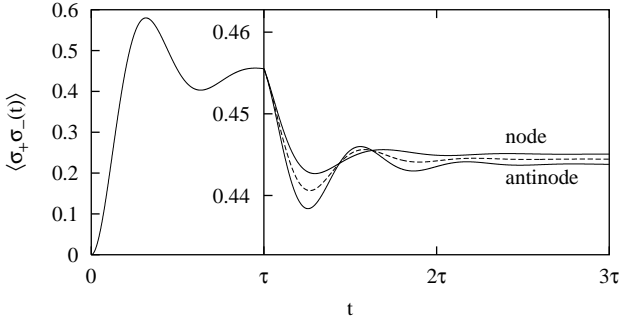}
 \caption{
   An example of the upper state population for $\Omega_0=2\gamma$,
   $\gamma\tau = 5$, $\varepsilon = 0.05$ and $\Delta=0$.  Until
   $t=\tau$ the behavior equals to that of free space.  In the
   magnified part ($t\ge\tau$) the functions are plotted for an atom
   in an antinode ($\omega_L\tau=2n\pi$), a node
   ($\omega_L\tau=(2n+1)\pi$) and for $\varepsilon=0$, i.e.\ no mirror
   (dashed line).
\label{fig:pop_high}}
\end{figure}
A numerically calculated example of the transient upper state
population is shown in Fig.~\ref{fig:pop_high}.

The steady state solution can be found by calculating the eigenvector
of the matrix $A_4 + \varepsilon K(\tau)$ with eigenvalue $0$. 
\begin{figure}[t]
  \centering \includegraphics[]{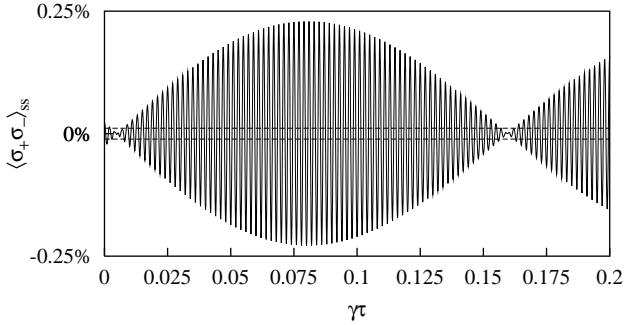}
 \caption{
   Excited state population deviation from the free space value
   ($\varepsilon=0$) in the long time limit depending on the distance
   between atom and mirror for $\Delta=0$, $\varepsilon=0.1$.  The
   laser intensity is taken to be rather strong ($\Omega_0 =
   20\gamma$). The rapid oscillations have to be regarded in a rather
   symbolic way; in a realistic situation the frequency would be much
   larger. Also indicated are the maximum and minimum values (dashed
   lines) of the oscillations obtained by
   Eq.~(\ref{eq:state_oscillation_near}).
\label{fig14}}
\end{figure}
From the form of the matrix $K(\tau)$ we expect that the laser
intensity (which is contained in the functions $f_j(\tau)$) influences
the decay rate(s) and driving force(s) in a steady state regime. In
fact, we see in Fig.~\ref{fig14} that the difference between the upper
state population obtained from Eq.~(\ref{eq:DDGL3}) and the results of
Sec.~\ref{sec:close} (indicated by the dashed lines) can be
significant. Furthermore, for $\Omega_0\gg\gamma$, small $\varepsilon$
and $\Delta=0$ the upper state population takes approximately the form
\begin{equation}
\langle\sigma_+\sigma_-\rangle_{ss} 
\approx \frac{\Omega_0^2}{\Gamma}
\bigg( 1+2\varepsilon\frac{\gamma^2}{\Gamma}\cos(\omega_0\tau) g(\tau) \bigg) 
\label{eq:upper_state_approx}
\end{equation}
with $\Gamma\equiv \gamma^2+2\Omega_0^2$. This expression equals 
Eq.~(\ref{eq:state_oscillation_near}) obtained in the Markovian limit
except for the function $g(\tau)$ which is given by
\begin{equation}
g(\tau) = \e^{-\frac{3}{4}\gamma\tau} 
\bigg( \frac{3}{4}\cos(\Omega_0\tau) - \frac{\Omega_0}{2\gamma}\sin(\Omega_0\tau)\bigg) 
+\frac{1}{4}\e^{-\frac{\gamma}{2}\tau}.
\end{equation}
We see that there is a modulation in the steady state population
defined by the Rabi frequency. This function has zero values in
regimes $\Omega_0\tau\approx n\pi$ independently of $\omega_0\tau$.
Thus, a strong laser can, in a way, inhibit the inhibition or
enhancement of spontaneous decay.

We will consider now the spectrum of the emitted light in
the channel parallel to the mirror.  For our purposes it turns out to
be advantageous to define an emission spectrum in terms of the mean
photon number increase $N(t)$ of that channel in the long time limit,
i.e.\ with the help of the differential equation (in a non-rotating frame)
\begin{equation}
\dot b_\omega = -\ii\omega b_\omega + \kappa_\omega\sigma_-,
        \quad\text{with}\quad\kappa_\omega\equiv\sqrt{\frac{2}{c}}h_k
\label{eq:b_dgl}
\end{equation}
we obtain
\begin{eqnarray}
\underset{t\rightarrow\infty}{\lim}\,\dot N(t) &=&
\underset{t\rightarrow\infty}{\lim}\,\frac{\partial}{\partial t}
             \int\dd\omega\,\langle b^\dagger_\omega(t) b_\omega(t)\rangle \nonumber\\
&=& (1-\varepsilon)\gamma\int\dd\omega\,S(\omega),
\end{eqnarray}
where we defined the spectrum
\begin{equation}
S(\omega) \equiv \frac{1}{\pi\kappa_\omega}\underset{t\rightarrow\infty}{\lim}\, 
         \text{Re}\big\{ \langle\sigma_+(t)b_\omega(t)\rangle \big\}.
\label{eq:spectrum}
\end{equation}
The usual expression including the Fourier transformed of an atomic
two-time correlation function is obtained (except for constant
factors) by integrating Eq.~(\ref{eq:b_dgl}) and inserting the result
in Eq.~(\ref{eq:spectrum}).  Furthermore, corresponding to an operator
$O$ we define its fluctuating part $\delta O \equiv O - \langle
O\rangle$ with the help of which we can split the spectrum in a
coherent and an incoherent component
\begin{equation}
S(\omega) = S_{\text{coh}}(\omega) + S_{\text{inc}}(\omega) \nonumber
\end{equation}
with
\begin{subequations}
\begin{eqnarray}
S_{\text{coh}}(\omega) &=& \frac{1}{\pi\kappa_\omega}\underset{t\rightarrow\infty}{\lim}\, 
         \text{Re}\big\{ \langle\sigma_+(t)\rangle\langle b_\omega(t)\rangle \big\},\\
S_{\text{inc}}(\omega) &=&  \frac{1}{\pi\kappa_\omega}\underset{t\rightarrow\infty}{\lim}\, 
         \text{Re}\big\{ \langle\delta\sigma_+(t)\delta b_\omega(t)\rangle \big\}.
\end{eqnarray}
\end{subequations}
It is easy to see that the coherent part of the spectrum takes the form 
\begin{equation}
S_{\text{coh}} = \langle\sigma_+\rangle_{ss}\langle\sigma_-\rangle_{ss}\delta(\omega-\omega_L),
\end{equation}
where the stationary values are taken from Eq.~(\ref{eq:DDGL3}).

In order to calculate the incoherent component of the spectrum we can
use a similar method as in the derivation of Eq.~(\ref{eq:DDGL3}).  It
is possible to derive a set of equations for the expectation values
\begin{equation}
\big( \langle\delta\sigma_-(t)\delta b_\omega(t)\rangle,\,
      \langle\delta\sigma_+(t)\delta b_\omega(t)\rangle,\,
      \langle\delta\sigma_z(t)\delta b_\omega(t)\rangle \big)^T \equiv \vec P(t)
\label{eq:vec_def}
\end{equation}
which takes in a rotating frame the form 
\begin{equation}
\dot{\vec P}(t) = \big[-\ii(\omega-\omega_L)\openone + A_3\big] 
                  \vec P(t) + \kappa_\omega\vec I_0(t) + \varepsilon \vec I(t,\tau).
\label{eq:spec_equation}
\end{equation}
Details of the calculation are given in Appendix~\ref{app:4}.
The last term in the above equation includes two-time correlation
functions which are again calculated in $0$-th order $\varepsilon$.
This yields in the long time limit an expression of the form
\begin{equation}
\vec P_{ss} = -\kappa_\omega M^{-1}
                           (\vec I_{0,ss} + \varepsilon \vec I_1(\tau))
\label{eq:spec_implicit}
\end{equation}
with
\begin{equation}
M = -\ii(\omega-\omega_L)\openone + A_3 +
                         \varepsilon \e^{-\ii(\omega-\omega_L)\tau} \tilde K(\tau).
\end{equation}
The atomic steady state expectation values which are contained in this
expression are given by the steady state solution of the delay
OBEs~(\ref{eq:DDGL3}). From this the spectrum $\vec S_{\text{inc}}$
can be calculated whereas for $\varepsilon = 0$ we get the usual
Mollow-spectrum \cite{mollow}.
\begin{figure}[t]
  \centering \includegraphics[]{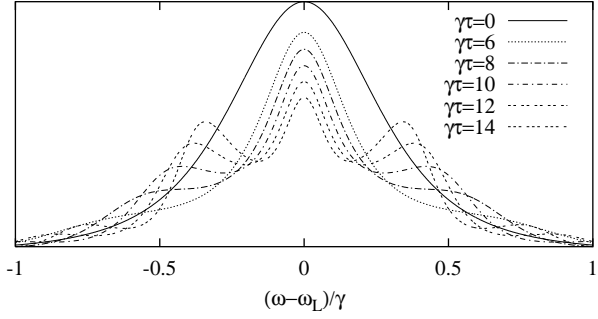}
 \caption{
   Incoherent emission spectra for various distances between atom and
   mirror and weak laser intensity ($\Omega_0= 0.2\gamma$, $\Delta=0$,
   $\varepsilon = 0.15$). The atom is always located in a node
   ($\omega_L\tau=2n\pi$).
\label{fig15}}
\end{figure}
\begin{figure}[t]
  \centering \includegraphics[]{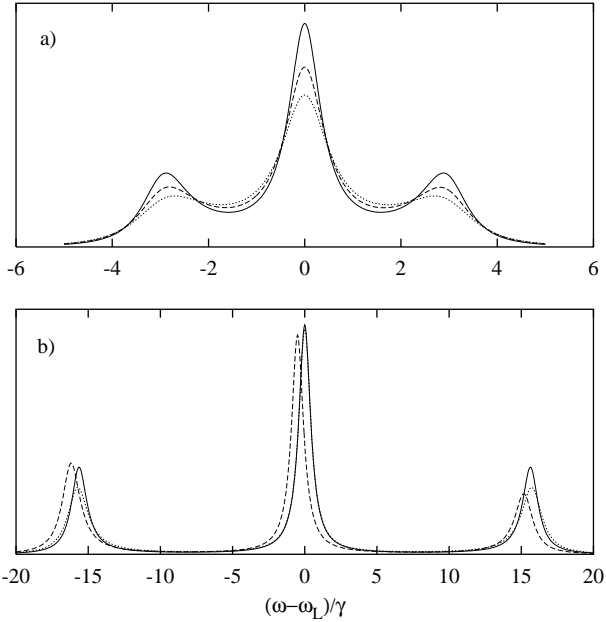}
 \caption{
   Emission spectra for an atom for higher laser intensities ($\Delta
   = 0$, $\varepsilon=0.2$).  In Fig. a) the quantity $\Gamma_0\tau$
   is small ($\Omega_0=3\gamma$). The atom is placed in a node (solid
   line, $\omega_L\tau=2n\pi$, $\gamma\tau=0.01$) at a slope (dashed
   line, $\omega_L\tau=(2n+\frac{1}{2})\pi$, $\gamma\tau=0.005$) and
   in an antinode (dotted line, $\omega_L\tau=(2n+1)\pi$,
   $\gamma\tau=0.02$).  In Fig. b) the laser intensity is higher
   ($\Omega_0=5\pi\gamma$). Solid line: Node position,
   $\gamma\tau=0.1$. Dashed line: Slope position,
   $\gamma\tau=0.1001$. For visibility, this line is horizontally
   displaced by a small amount. Dotted line: Antinode position
   $\gamma\tau=0.1002$.
  \label{fig16}}
\end{figure}
\begin{figure}[ht]
  \centering \includegraphics[]{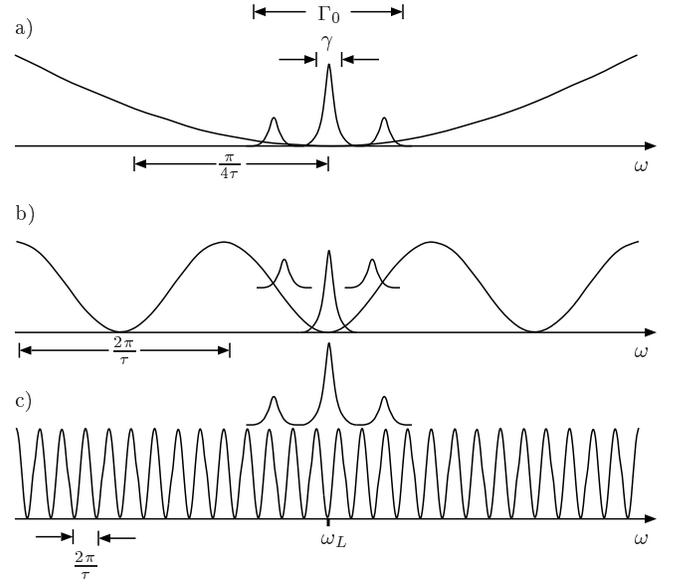}
 \caption{
   Illustration of the explanation for the different shapes of the
   emission spectra (see text) for increasing values of
   $\Gamma_0\tau$.  The curves in the figures correspond to the
   function $\sin^2(\omega\tau/2)$ for different values of $\tau$. For
   larger atom-mirror distances the relative variation of this
   function in a region $\Gamma_0$ (the width of the fluorescence
   triplet) becomes significant and the Markov approximation is not
   valid anymore.  The atom is placed at a node in this example.
\label{fig17}}
\end{figure}

Examples obtained from Eq.~(\ref{eq:spec_implicit}) are shown in
Fig.~\ref{fig15} for weak laser intensity and an atom at a node.
The spectrum for $\gamma\tau = 0$ in this figure is the
Mollow result with a damping rate $(1-\varepsilon)\gamma$.  The
structures arising at large distances resemble those of
Fig.~\ref{fig6} and can be interpreted in a similar way.  The
situation changes in case of higher laser intensities.  Examples
are shown in Fig.~\ref{fig16}a) and \ref{fig16}b) for different values
of $\tau$ and different positions of the atom.  We see that in
general the width of the spectra varies and they are asymmetric
depending on the position of the atom.

This behavior can be understood at least on a qualitative level if we
take into account that a measure of the coupling strength of the atom
to a field mode of frequency $\omega$ is given by
$\sin^2(\omega\tau/2)$.  This function varies in frequency space on a
scale $1/\tau$. Defining the quantity $\Gamma_0\equiv 2\Omega+\gamma$
which approximately gives the overall width of the triplet we see that
for $\Gamma_0\tau\ll1$ (and which we take now as the criterion for
small atom-mirror distance) that the coupling is almost flat in the
region where the spectrum differs from zero (see Fig.~\ref{fig17}a)).
This situation corresponds to the Markovian limit discussed in
Sec.~\ref{sec:close}. Thus, we obtain in good approximation the usual
Mollow spectrum with a modified spontaneous emission rate
$\tilde\gamma_L$. This is shown in Fig.~\ref{fig16}a) for various
positions of the atom. The level shift (\ref{eq:detuning_shift}),
which acts here as a detuning in case of the dashed line, is so
small that this curve cannot be distinguished from the Mollow spectrum
with decay rate $\gamma$ on the scale of the figure.  For larger
values of $\Gamma_0\tau$, but still $\gamma\tau\ll1$ , we have a
situation like it is shown in Fig.~\ref{fig17}b) where, as an example,
an atom located at a node of the standing wave $\sin^2(\omega\tau/2)$
is chosen. For increasing laser intensity, the sidebands move towards
regions of higher values of the coupling function leading to a higher
damping of, say, the corresponding levels in a dressed state picture
and thus to a broadening of the sidebands (with increasing laser
intensity until $\Omega_0\approx\pi/\tau$). For an atom placed at an
antinode the behavior is simply the inverse.  However, if the atom is
placed at a ``slope'', e.g. the one on the right hand side of the node
which was considered in Fig.~\ref{fig17}b), the spectrum becomes
asymmetric since the transition responsible for the right sideband is
stronger damped than the left one. Thus, the right sideband is broader
than the left sideband which is in accordance with the dashed line in
Fig.~\ref{fig16}b) (For the sake of clarity, the dashed line in this figure is
displaced horizontally by a small amount). The case
$\Gamma_0\tau\gg1$ is indicated in Fig.~\ref{fig17}c) leading to
structures as in Fig.~\ref{fig15} or Fig.~\ref{fig6}.

\begin{figure}[t]
  \centering \includegraphics[]{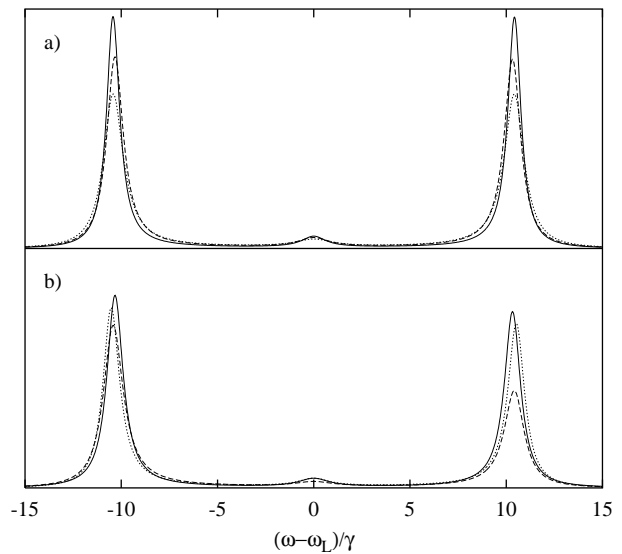}
 \caption{
   Incoherent emission spectra for non-vanishing laser detuning
   ($\Delta=10\gamma$, $\varepsilon=0.2$, $\Omega_0=3\gamma$). Fig.~a)
   shows an atom very close to the mirror: The solid line corresponds
   to node positions ($\omega_L\tau=2n\pi$, $\gamma\tau=0.002$), the
   dashed line to a slope position
   ($\omega_L\tau=(2n+\frac{1}{2})\pi$, $\gamma\tau=0.0025$) and the
   dotted line to an antinode position ($\omega_L\tau=(2n+1)\pi$,
   $\gamma\tau=0.001$).  In Fig.~b) $\Gamma_0\tau$ is larger. Solid
   line: $\omega_L\tau=2n\pi$, $\gamma\tau=0.15$, dashed line:
   $\omega_L\tau=(2n+\frac{1}{2})\pi$, $\gamma\tau=0.1505$, dotted
   line: $\omega_L\tau=(2n+1)\pi$, $\gamma\tau=0.151$.
\label{fig18}}
\end{figure}

So far we have discussed the case of exact resonance ($\Delta=0$)
where the emission spectra are symmetric for an atom in a node or an
antinode. This situation changes, in general, if we take a finite
laser detuning.  In case of $\Gamma_0\tau\ll1$ the spectra are
approximately identical to the usual Mollow spectra with modified
spontaneous emission rate $\tilde\gamma_L$ and detuning
$\tilde\Delta$, i.e.\ they are approximately symmetric independent of
the exact atomic position. Examples for this case are shown in
Fig.~\ref{fig18}a). Note that the sideband positions for an atom
located at a slope are shifted towards the central peak which is due
to the small frequency shift (the sideband positions are approximately
given by $\omega_L\pm\sqrt{\Omega_0^2 +
  (\Delta-\varepsilon\gamma/2\sin(\omega_L\tau))^2}$).
This situation differs from that when the distance between the atom
and the mirror is increased. Here, the spectra become asymmetric even
when the atom is located in a node or an antinode (see
Fig.~\ref{fig18}b)).

\section{Summary\label{sec:summary}}
In this work we have discussed the behavior of an atom in the presence of a
reflecting wall with regard to pure spontaneous emission, i.e.\ the
decay of an initially excited atom without any laser excitation, and
with regard to an additional continuous driving laser field. In the
first case, the one dimensional model applied here, can be solved
exactly leading to a solution which directly reveals the 
retarded character of the system (photons bouncing back an forth
between the atom and the mirror) visible in the state population, the
field intensity and the (transient) photon spectrum.  The limit of
small distances yields the usual behavior of enhanced and
inhibited spontaneous emission which can be interpreted as an
interference phenomenon of the the outgoing and reflected light pulse
leading to a standing wave pattern in the field intensity: If the atom
is placed in an antinode of this standing wave, spontaneous decay is
enhanced while in a node it is suppressed.  For large distances this
interference is not significant anymore and the node-antinode location
of the atom becomes less important. The emitted photon wave packet is
back reflected by the mirror leading to a partial re-excitation of the
atom which starts now to emit again radiation and so on.

In case of an additional driving laser the situation is more complex
since the energy of the system increases continuously. Working in a
Heisenberg picture we have derived a set of equations which
serves as a starting point for several approximative treatments.  In
the limit of low laser intensities we saw with the help of
perturbation theory and a harmonic oscillator model that the system
behaves essentially like an atom driven by two monochromatic lasers
where the phase difference between the lasers is controlled by the
atom-mirror distance. The intensity of the reflected light at the
position of the atom depends on the intensity of the driving force on
the atom at a preceding time which leads in general to a different
state population in every time interval $[m\tau,(m+1)\tau]$
(converging to a steady state value).  The dominance of coherent
scattering was confirmed by the monochromatic emission spectrum of the
system.  In this limit we gave also a discussion of the second order
intensity correlation function which included in case of the field in
channel $1$ an interference of different paths leading to a
coincidence signal. This fact causes non-trivial structures in the
correlation function. 

In case of a higher laser intensity incoherent scattering becomes more
significant.  However, for small solid angles $\varepsilon$ it is
possible to derive a closed set of linear delay differential equations
which represents an extension of the usual OBEs.  It turned out that an
intense laser field can significantly influence the system if we
compare it with the Markovian limit where it is possible to describe
the system by OBEs with modified decay rate and
transition frequency. With regard to the upper state population, for
instance, the laser can make the effect of the mirror to disappear
regardless of the exact position of the atom (node or antinode).
Furthermore, the influence of a strong laser was revealed by the
emission spectra.  Even if the width of the three peaks of the
spectrum are each very small compared to the inverse delay time
$\tau$, an intense laser field (or a high detuning) can ``push'' the
sidebands of the triplet towards regions of a higher or lower coupling
of the corresponding transitions to the radiation field leading to
features like asymmetric spectra (see also \cite{mossberg1994}).

A possible extension of the discussion presented in this article would
be the inclusion of the motional degrees of freedom of the atom.
Assuming an atom in a harmonic trap, as in the experimental
realization \cite{eschner}, the reflected radiation will have an
appreciable effect on the center of mass motion of the ion. This can
serve as a further probe for effects discussed in this paper.  Besides
that, collective effects of two ions in the trap, like super- and
sub-radiance, could be studied when the image of one ion is projected
onto the other.  The effect of one atom on another one, mediated by
radiation over a large distance, is important for applications like
quantum communication \cite{vanenk}.

\begin{acknowledgments}
  We thank J. Eschner, P. Bouchev and R. Blatt for stimulating and helpful
  discussions.  This work was supported by the Austrian Science
  Foundation and the European Union TMR network Cold Quantum Gases
  (Contract No: HPRN-CT-2000-00125).
\end{acknowledgments}

\appendix
\section{The scattered light field\label{app:1}}
Here we will sketch the derivation of an
expression for the field intensity in the channel perpendicular to the
mirror which was used for the generation of Fig.~\ref{fig3} and
Fig.~\ref{fig5}.  Besides that, we will
give formulae for the electric field operators in the Heisenberg
picture which are used in the discussion of laser excitation and an
expression for the second order correlation function which is needed
to derive Eq.~(\ref{eq:G_1(t)}). We use the
coordinate system introduced in Fig.~\ref{fig1}.

The intensity of the emitted light corresponding to
Sec.~\ref{sec:sec1} is defined by
\begin{eqnarray}
I(z,t) &=& \langle E^\dagger(z,t)E(z,t)\rangle 
        = \langle\psi(t)\vert E^\dagger(z)E(z) \vert \psi(t)\rangle \nonumber\\
&=& \left\vert \ii\int \dd k\,
     \alpha_k\sin(kz)b_{g,k}^1(t) \right\vert^2 \equiv \vert A(z,t)\vert^2.
\label{eq:int_def}
\end{eqnarray}
From Eq.~(\ref{eq:dgl0}) we see that
\begin{align}
  A(z,t) =
  &-\varepsilon\frac{\gamma}{2}\frac{\ii\hbar}{d}\frac{1}{2\pi}
  \int_0^t\dd t'\,\e^{-\ii\omega_0t'}b_e(t') \nonumber\\
  \times&\int\dd \omega\,\e^{\ii\omega(t'-t)} \left(
    \e^{\ii\omega\frac{\tau}{2}}-\e^{-\ii\omega\frac{\tau}{2}} \right)
  \left( \e^{\ii\omega \frac{z}{c}}-\e^{-\ii\omega \frac{z}{c}}
  \right),
\end{align}
where the frequency integral gives rise to delta functions which yield
non-vanishing terms only in certain regions of space and time,
\begin{align}
  A(z,t) =& \varepsilon\frac{\gamma}{2}\frac{\ii\hbar}{d}\e^{-\ii\omega_0t} \nonumber\\
  \times\bigg( &\e^{\ii\omega_0(\frac{z}{c}-\frac{\tau}{2})}
  b_e(t-{\textstyle\frac{z}{c}}+{\textstyle\frac{\tau}{2}})
  \Theta(t-{\textstyle\frac{z}{c}}+{\textstyle\frac{\tau}{2}})
  \Theta({\textstyle\frac{z}{c}}-{\textstyle\frac{\tau}{2}}) \nonumber\\
  +&\e^{-\ii\omega_0(\frac{z}{c}-\frac{\tau}{2})}
  b_e(t+{\textstyle\frac{z}{c}}-{\textstyle\frac{\tau}{2}})
  \Theta(t+{\textstyle\frac{z}{c}}-{\textstyle\frac{\tau}{2}})
  \Theta({\textstyle\frac{\tau}{2}}-{\textstyle\frac{z}{c}}) \nonumber\\
  -&\e^{\ii\omega_0(\frac{z}{c}+\frac{\tau}{2})}
  b_e(t-{\textstyle{\textstyle\frac{z}{c}}}-{\textstyle\frac{\tau}{2}})
  \Theta(t-{\textstyle\frac{z}{c}}-{\textstyle\frac{\tau}{2}}) \bigg).
\label{eq:intensity1}
\end{align}
The first and the second line of this expression represent the
outgoing light pulses to the right and the left side, respectively
while the last line provides us with the reflected light pulse.

In case of laser excitation the appropriate quantity in order to
calculate the intensity is the electric field operator in the
Heisenberg picture. Starting with the Heisenberg equations
(\ref{eq:heis1}) and the definitions~(\ref{eq:field_op}), we use a
similar derivation as above, where $A(z,t)$ has to be replaced by the
operator $E_1(z,t)$ and the amplitude $b_e$ by $\sigma_-$ (and
$\omega_0$ by $\omega_L$ for a detuned laser). Apart from an
additional noise term the result coincides with
Eq.~(\ref{eq:intensity1}). On condition that a photo detector is
placed on the right hand side of the atom (cf.  Fig.~\ref{fig11}) at a
position $z_0>L$ the second line in Eq.~(\ref{eq:intensity1}) does not
contribute and one gets
\begin{align}
  E_1(d_0,t) =
  \varepsilon\frac{\gamma}{2}\frac{\ii\hbar}{d}&\e^{-\ii\omega_L(t-\frac{d_0}{c})}
  \bigg( \sigma_-(t-{\textstyle\frac{d_0}{c}})
  \Theta(t-{\textstyle\frac{d_0}{c}}) \nonumber\\
  &-\e^{\ii\omega_L\tau} \sigma_-(t-{\textstyle\frac{d_0}{c}}-\tau)
  \Theta(t-{\textstyle\frac{d_0}{c}}-\tau )
  \bigg)\nonumber\\
  &+N_1(d_0,t)
\label{eq:field1}
\end{align}
with
\begin{equation}
N_1(d_0,t) =  \frac{\ii\hbar}{d}\int \dd k\,g_k
                     \sin(k(d_0+L))a_k(0)\e^{-\ii\omega_k t},
\end{equation}
where $d_0=z_0-L=z_0-c\tau/2$ is the distance between the detector and
the atom. There are two different kinds of signals arriving at the
detector, one which takes its way directly and one which takes the
``loop way'' over the mirror and therefore needs a longer time.

If the conditions of Sec.~\ref{sec:close} are fulfilled, we can
approximately calculate the intensity in channel $1$ by neglecting
$\tau$ in the arguments of the operators and the step function to
obtain
\begin{align}
\langle E_1^\dagger(d_0,t) E_1(d_0,t)\rangle =&  
\left( \frac{\varepsilon\gamma\hbar}{d}\right)^2
  \sin^2(\omega_L\tau/2) \nonumber\\
&\times\langle\sigma_+(t-{\textstyle\frac{d_0}{c}})
                                \sigma_-(t-{\textstyle\frac{d_0}{c}})\rangle
\end{align}
which leads to expression (\ref{eq:intensity0}) if $d_0$ is set to
zero.

For the sake of completeness we give here also the electric field in
channel 2 since it is used for various calculations,
\begin{align}
  &E_2(x,t) = \frac{(1-\varepsilon)\gamma}{2}\frac{\ii\hbar}{d}\e^{-\ii\omega_L t} \nonumber\\
  &\times\bigg( \e^{\ii\frac{x}{c}}\sigma_-(t-{\textstyle\frac{x}{c}})
  \Theta(t-{\textstyle\frac{x}{c}})\Theta({\textstyle\frac{x}{c}}) \nonumber\\
  &\quad +\e^{-\ii\frac{x}{c}}\sigma_-(t+{\textstyle\frac{x}{c}})
  \Theta(t+{\textstyle\frac{x}{c}})\Theta(-{\textstyle\frac{x}{c}})\bigg)
  + N_2(x,t)
\label{eq:field2}
\end{align}
with
\begin{equation}
N_2(x,t) = \frac{\ii\hbar}{d}\int \dd k\,h_kb_k(0)\e^{\ii(kx-\omega_kt)}.
\end{equation} 
With the help of Eq.~(\ref{eq:field1}) one can easily find expressions
for the intensity and the first order field correlation function in
channel $1$ (the functions connected with the channel $2$ coincide
with those of standard Markovian theory).

Using Eq.~(\ref{eq:field1}) and the commutation
relations~(\ref{eq:commutator}) we get also an expression for the
second order correlation function~(\ref{eq:G2}),
\begin{align}
  G^{(2)}_1(t,t+T) = \parallel [&
  \sigma_-(t+T)\sigma_-(t) \nonumber\\
  +& \sigma_-(t+T-\tau)\sigma_-(t-\tau) \nonumber\\
  -& \e^{\ii\omega_L\tau}\sigma_-(t+T)\sigma_-(t-\tau) \nonumber\\
  -&
  \e^{\ii\omega_L\tau}\mathcal{T}_{_\leftarrow}\sigma_-(t+T-\tau)\sigma_-(t)
  ]\,\vert G\rangle\parallel^2.
\label{eq:G_1^2}
\end{align}
We set the arbitrary distance $d_0$ to zero and omitted the step
functions in this expression which is valid if $t>\tau$ (if not,
components with negative arguments are simply set to zero). The effect
of the non vanishing commutator in Eq.~(\ref{eq:commutator}) is to
conserve time ordering in the last term of Eq.~(\ref{eq:G_1^2}), i.e.\
the time argument of the operator on the left hand side is always
greater than the right one.  This is indicated by the symbol
$\mathcal{T}_{_\leftarrow}$. We see that for $T<\tau$ and $T>\tau$ the
operators have to be exchanged.

\section{Some features of the harmonic oscillator model\label{app:2}}
In order to derive an expression for the correlation functions in the
harmonic oscillator model we start with the Heisenberg equation of
motion for the operator $c$,
\begin{align}
  \frac{\dd}{\dd t}c(t) =
  &-\alpha_1 c(t) + \alpha_2 c(t-\tau)\Theta(t-\tau) + \alpha_3 \nonumber\\
  &+ \frac{\ii d}{\hbar}\e^{\ii\omega_L t}\big( N_1(t) + N_2(t) \big).
\label{eq:Heisi}
\end{align}
with parameters $\alpha_i$ are defined in Sec.~\ref{sec:low_laser} and noise
operators given by (\ref{eq:noisy}). Let us define a vector
\begin{equation}
\vert\Psi(t)\rangle \equiv c(t)\vert\psi(0)\rangle,
\quad\text{i.e.}\quad\vert\Psi(0)\rangle=c(0)\vert\psi(0)\rangle,
\end{equation}
where $\vert\psi(0)\rangle \equiv \vert\varphi,\{0\}_1,\{0\}_2\rangle$ is the
initial state of the system. The state
\begin{equation}
\vert\varphi\rangle = a \vert g\rangle + b \vert e\rangle,
\quad \vert a\vert^2 + \vert b\vert^2 = 1
\end{equation}
is an arbitrary state on the atomic space.
Eq.~(\ref{eq:Heisi}) provides us with an equation of motion for this
vector,
\begin{align}
  \frac{\dd}{\dd t}\vert\Psi(t)\rangle =
  &-\alpha_1 \vert\Psi(t)\rangle + \alpha_2 \vert\Psi(t-\tau)\rangle\Theta(t-\tau) 
\nonumber\\
  &+ \alpha_3\vert\psi(0)\rangle.
\label{eq:oscillator_equation}
\end{align}
Thus we have a linear inhomogeneous delay differential equation and
its solution takes the form
\begin{equation}
\vert\Psi(t)\rangle = \text{A}(t)\vert\psi(0)\rangle 
                    + \text{B}(t)\vert\Psi(0)\rangle. 
\end{equation}
This vector has the form
$\vert\chi(t)\rangle\vert\{0\}_1,\{0\}_2\rangle$ where the
non-constant part is an element of the atomic Hilbert space. The
coefficients are given by
\begin{align}
  \text{A}(t) &= \frac{\alpha_3}{\alpha_1}
  \sum_{n=0}^\infty \frac{\alpha_2^n}{n!}(t-n\tau)^n G_n[-\alpha_1(t-n\tau)]\Theta(t-\tau), \\
  \text{B}(t) &= \sum_{n=0}^\infty \frac{\alpha_2^n}{n!}(t-n\tau)^n
  \e^{-\alpha_1(t-n\tau)}\Theta(t-n\tau).
\end{align}
These expressions can be found by Laplace transformation in a way it
was demonstrated in Eq.~(\ref{eq:demo}) (the function $G_n$ is defined
in Eq.~(\ref{eq:Gn})). 

Some expectation values of interest are
\begin{align}
&\langle c(t)\rangle = \langle\psi(0)\vert\Psi(t)\rangle =
  \text{A}(t) + \text{B}(t)\langle c(0)\rangle,
\label{eq:dipole}\\
&\langle c^\dagger(t) c(t)\rangle
= \|\vert\Psi(t)\rangle\|^2 \nonumber\\
&\quad= \langle c^\dagger(t)\rangle\langle c(t)\rangle 
+ \vert\text{B}(t)\vert^2 \big(\langle c^\dagger(0) c(0)\rangle -
\langle c^\dagger(0)\rangle\langle c(0)\rangle\big).
\label{eq:upper_pop} 
\end{align}
From the above expressions it is immediately clear that
\begin{equation}
\langle
c^\dagger(t) c(t)\rangle 
  = \langle c^\dagger(t)\rangle\langle c(t)\rangle\quad\forall\, t>0
\label{eq:upper_pop_fac}
\end{equation}
if $\langle c^\dagger(0) c(0)\rangle = \langle c^\dagger(0)\rangle\langle
c(0)\rangle$, which is the case iff the atom is initially in the
ground state. In the long time limit this behavior is independent of
the initial state, i.e.\ $\langle c^\dagger c\rangle_{ss} = \langle
c^\dagger\rangle_{ss}\langle c\rangle_{ss}$ since
$\underset{t\rightarrow\infty}{\lim}\text{B}(t) = 0$.

If we start in the ground state the harmonic oscillator model
reproduces result (\ref{eq:perturb_result}) gained from the
perturbation theory and a remarkable fact is that
Eq.~(\ref{eq:upper_pop}) also reproduces the solution (\ref{eq:sol1})
obtained from the modified Wigner-Weisskopf theory if we set
$\alpha_3=0$ (no laser) so that $\text{A}(t)=0$ for all $t$ and
$\vert\varphi\rangle=\vert e\rangle$.

The two-time correlation functions take the form
\begin{align}
&\langle c^\dagger(t')c(t) \rangle
  = \langle\Psi(t')\vert\Psi(t) \rangle \nonumber\\
&\quad=  \langle c^\dagger(t')\rangle\langle c(t)\rangle 
  + \text{B}^*(t')\text{B}(t) \big(\langle c^\dagger c(0)\rangle
  - \langle c^\dagger(0)\rangle\langle c(0)\rangle\big),
\label{eq:upper_cor}
\end{align}
where the time order is irrelevant. We see again that this quantity
factorizes in special cases,
\begin{eqnarray}
&&\langle c^\dagger(t')c(t) \rangle 
= \langle c^\dagger(t')\rangle\langle c(t) \rangle,
    \quad\text{if}\,\vert\varphi\rangle=\vert g\rangle, 
\label{eq:upper_cor_fac}\nonumber\\
&&    \underset{t\rightarrow\infty}{\lim}\langle c^\dagger(t+T)c(t) \rangle 
= \langle c^\dagger\rangle_{ss}\langle c\rangle_{ss},
\quad\forall\,\vert\varphi\rangle.
\label{eq:steady_fac2}
\end{eqnarray}

In order to derive fourth order correlation functions we proceed in a
similar way. We define a vector
\begin{eqnarray}
\vert\Phi(t')\rangle &\equiv& c(t')\vert\Psi(t)\rangle, \\
\vert\Phi(0)\rangle&=&c(0)\vert\Psi(t)\rangle=\text{A}(t)\vert\Psi(0)\rangle.
\end{eqnarray}
For this vector we get again an equation of the
form~(\ref{eq:oscillator_equation}) where we replace
$\vert\Psi(t)\rangle\rightarrow\vert\Phi(t')\rangle$ and
$\vert\psi(0)\rangle\rightarrow\vert\Psi(t)\rangle$ which finally gives the fourth
order correlation function
\begin{eqnarray}
G^{(2)}_2(t,t+T) 
&=& \langle c^\dagger(t)c^\dagger(t+T)c(t+T)c(t) \rangle \nonumber\\
&=& \|\vert\Phi(t+T)\rangle\|^2 
\end{eqnarray}
which leads to
\begin{align}
  G^{(2)}_2(t,t+T) =
  \| \big(&A(t)A(t+T)+A(t+T)B(t)c(0) \nonumber\\
  + &A(t)B(t+T)c(0)\big)\vert\psi(0)\rangle \|^2,
\end{align}
and thus,
\begin{align}
  &G^{(2)}_2(t,t+T) = \langle c^\dagger(t) c(t)\rangle
  \langle c^\dagger(t+T)c(t+T)\rangle,
  \,\vert\varphi\rangle=\vert g\rangle, \nonumber\\
  &\underset{t\rightarrow\infty}{\lim} G^{(2)}_2(t,t+T) = \langle
  c^\dagger c\rangle_{ss}^2, \quad\forall\,\vert\varphi\rangle.
\end{align}
From Eq.~(\ref{eq:G_1^2}) we also obtain
\begin{equation}
 \underset{t\rightarrow\infty}{\lim} G^{(2)}_1(t,t+T) 
= 16\sin^4(\omega_L\tau/2)\langle c^\dagger c\rangle_{ss}^2,
 \quad\forall\,\vert\varphi\rangle.
\end{equation}
This result is equal to the square of the intensity in the long time
limit in this channel.

\section{Calculation of the spectrum\label{app:4}}
In this appendix a sketch of the derivation of the emission spectrum
in case of a higher laser intensity is outlined. We will indicate in
the following merely retarded time arguments. In order to get
Eq.~(\ref{eq:spec_equation}) and Eq.~(\ref{eq:spec_implicit}) we
consider the operators
\begin{eqnarray}
\delta A&\equiv&\delta\sigma_-\delta b_\omega =   \sigma_-b_\omega 
                             + \langle\sigma_-\rangle\langle b_\omega\rangle 
                             - b_\omega\langle\sigma_-\rangle
                             - \sigma_-\langle b_\omega\rangle, \nonumber\\
\delta B&\equiv&\delta\sigma_+\delta b_\omega = \sigma_+b_\omega 
                             + \langle\sigma_+\rangle\langle b_\omega\rangle 
                             - b_\omega\langle\sigma_+\rangle
                             - \sigma_+\langle b_\omega\rangle, \nonumber\\
\delta C&\equiv&\delta\sigma_z\delta b_\omega = \sigma_zb_\omega 
                             + \langle\sigma_z\rangle\langle b_\omega\rangle 
                             - b_\omega\langle\sigma_z\rangle
                             - \sigma_z\langle b_\omega\rangle. \nonumber\\
&&\label{eq:delta_operators}
\end{eqnarray}
After transforming in a rotating frame, $\delta A \rightarrow
\e^{-2\ii\omega_Lt}\delta A$, $\delta B\rightarrow \delta B$, $\delta
C \rightarrow \e^{-\ii\omega_Lt}\delta C$, $\sigma_- \rightarrow
\e^{-\ii\omega_Lt}\sigma_-$, $b_\omega\rightarrow \e^{-\ii\omega t}
b_\omega$, the Heisenberg equations of motion for these operators
yield, after taking the expectation value,
Eq.~(\ref{eq:spec_equation}),
\begin{equation}
\dot{\vec P}(t) = \big[-\ii(\omega-\omega_0)\openone + A_3\big] 
                  \vec P(t) + \kappa_\omega\vec I_0(t) + \varepsilon \vec I(t,\tau),
\end{equation}
where $\vec P(t)$ is defined in Eq.~(\ref{eq:vec_def}) and
\begin{eqnarray}
A_3 &\equiv& \begin{pmatrix} -\frac{\gamma}{2}-\ii\Delta  & 0 & -\ii\frac{\Omega_0}{2} \\ 
                0 & -\frac{\gamma}{2}+\ii\Delta & \ii\frac{\Omega_0}{2} \\
                -\ii\Omega_0 & \ii\Omega_0 & -\gamma  
\end{pmatrix},\\
\vec I_0(t) &=& 
\begin{pmatrix}
  -\langle\sigma_-\rangle^2 \\
  \langle\sigma_+\sigma_-\rangle - \langle\sigma_+\rangle\langle\sigma_-\rangle \\
  -2\langle\sigma_+\sigma_-\rangle\langle\sigma_-\rangle
\end{pmatrix}.
\end{eqnarray}
The components of $I(t,\tau)$ are given by
\begin{align} 
  I_1(t,\tau) =& -\frac{\gamma}{2}\e^{\ii\omega_L\tau}\big(\langle
  \delta C\sigma_-(t-\tau)\rangle
  + \langle\sigma_z\rangle h_-(t,\tau)\big) \nonumber\\
  I_2(t,\tau) =& -\frac{\gamma}{2}\e^{-\ii\omega_L\tau}\big(\langle
  \sigma_+(t-\tau)\delta C\rangle
  + \langle\sigma_z\rangle h_+(t,\tau)\big) \nonumber\\
  I_3(t,\tau) =&\quad \gamma\e^{-\ii\omega_L\tau} \big(
  \langle\sigma_+(t-\tau)\delta A\rangle +
  \langle\sigma_-\rangle h_+(t,\tau) \big) \nonumber\\
  &+ \gamma\e^{\ii\omega_L\tau} \big(\langle\delta
  B\sigma_-(t-\tau)\rangle + \langle\sigma_+\rangle h_-(t,\tau)\big)
\label{eq:corr_term}
\end{align}
with
\begin{eqnarray}
h_-(t,\tau) \equiv \langle \delta b_\omega\sigma_-(t-\tau)\rangle\e^{\ii(\omega_L-\omega)t}, \\
h_+(t,\tau) \equiv \langle \sigma_+(t-\tau)\delta b_\omega\rangle\e^{\ii(\omega_L-\omega)t},
\label{eq:h_terms}
\end{eqnarray}
where only retarded time arguments are indicated. The quantities
$h_\pm$ can be calculated with the help of the Heisenberg equations of
motion for $b_\omega$. The result contains atomic two time correlation
functions which have to be calculated again in $0$-th order
$\varepsilon$ depending on the initial state which is a single time
expectation value.  The calculation of the correlation functions
contained in~(\ref{eq:corr_term}) is analogous to the calculation done
to derive the delay OBEs~(\ref{eq:DDGL3}): We multiply the Heisenberg
equations of motion for the operators~(\ref{eq:delta_operators}) once
from the left with $\sigma_+(t')$ and once from the right with
$\sigma_-(t')$ ($t'\le t$) and keep only terms of first order in
$\varepsilon$. The six equations we get in this way now contain again
atomic two time correlation functions which have to be calculated as
it was done for $h_\pm$.  Then we let $t\rightarrow\infty$ and get
an expression of the form~(\ref{eq:spec_implicit})
with
\begin{equation}
\tilde K(\tau) =
\begin{pmatrix} 
  \frac{\gamma}{2}\tilde f_1(\tau) & 0 & -\ii\frac{\Omega_0}{2}\tilde
  f_2(\tau)
  \\
  0 & \frac{\gamma}{2}\tilde f_1^*(\tau) & \ii\frac{\Omega_0}{2}\tilde
  f_2^*(\tau)
  \\
  - \ii\Omega_0\tilde f_3(\tau) & \ii\Omega_0\tilde f_3^*(\tau) &
  \gamma \tilde f_4(\tau)
\end{pmatrix},
\end{equation}
\begin{equation}
\tilde f_1(\tau) = -\e^{\ii\omega_L\tau} g_z(\tau), 
\end{equation}
\begin{equation}
\tilde f_2(\tau) = -\frac{\ii\gamma}{2\Omega_0}\e^{\ii\omega_L\tau}U_{31}^*(\tau), 
\end{equation}
\begin{equation}
\tilde f_3(\tau) = \frac{\ii\gamma}{\Omega_0}\e^{\ii\omega_L\tau}g_+(\tau), 
\end{equation}
\begin{equation}
\tilde f_4(\tau) = \frac{1}{2}
\big(\e^{-\ii\omega_L\tau}U_{11}(\tau) + \e^{\ii\omega_L\tau}U_{11}^*(\tau)\big), 
\end{equation}
and
\begin{equation}
\begin{pmatrix} 
  g_-(\tau) \\ g_+(\tau) \\ g_z(\tau)
\end{pmatrix} = 
U(\tau)\left[
\begin{pmatrix} 
  0 \\ 0 \\ -1
\end{pmatrix} -
\begin{pmatrix} 
  \langle\sigma_-\rangle_{ss} \\ \langle\sigma_+\rangle_{ss} \\
  \langle\sigma_z\rangle_{ss}
\end{pmatrix}
\right]
+
\begin{pmatrix} 
  \langle\sigma_-\rangle_{ss} \\ \langle\sigma_+\rangle_{ss} \\
  \langle\sigma_z\rangle_{ss}
\end{pmatrix}.
\end{equation}
The inhomogeneity $I_1(\tau)$ in Eq.~(\ref{eq:spec_implicit}) is so
lengthy that we do not quote it here.

\end{document}